\documentclass{svjour3}                     
\smartqed  
\usepackage{natbib}
\usepackage{graphicx}
\usepackage{amsmath}
\usepackage{amssymb}
 \newcommand{\Id}{{\tens I}}
 \newcommand{\di}{day$^{-1}$}
 \newcommand{\err}[3]{\left(#1,{+#2\atop-#3}\right)}
 \journalname{Celestial Mechanics \& Dynamical Astronomy}
\begin{document}

\title{Orbital structure of the GJ876 extrasolar planetary system, based on the latest
Keck and HARPS radial velocity data
}

\titlerunning{GJ876 orbital structure}        

\author{Roman V. Baluev}


\institute{
Pulkovo Astronomical Observatory, Pulkovskoje sh. 65/1, Saint Petersburg 196140, Russia; \\
Sobolev Astronomical Institute, Saint Petersburg State University,
Universitetskij pr. 28, Petrodvorets, Saint Petersburg 198504, Russia \\
           \email{roman@astro.spbu.ru}
}

\date{Received: date / Accepted: date}

\maketitle

\begin{abstract}
We use full available array of radial velocity data, including recently published HARPS
and Keck observatory sets, to characterize the orbital configuration of the planetary
system orbiting GJ876. First, we propose and describe in detail a fast method to fit
perturbed orbital configuration, based on the integration of the sensitivity equations
inferred by the equations of the original $N$-body problem. Further, we find that it is
unsatisfactory to treat the available radial velocity data for GJ876 in the traditional
white noise model, because the actual noise appears autocorrelated (and demonstrates
non-white frequency spectrum). The time scale of this correlation is about a few days, and
the contribution of the correlated noise is about $2$~m/s (i.e., similar to the level of
internal errors in the Keck data). We propose a variation of the maximum-likelihood
algorithm to estimate the orbital configuration of the system, taking into account the red
noise effects. We show, in particular, that the non-zero orbital eccentricity of the
innermost planet \emph{d}, obtained in previous studies, is likely a result of
misinterpreted red noise in the data. In addition to offsets in some orbital parameters,
the red noise also makes the fit uncertainties systematically underestimated (while they
are treated in the traditional white noise model). Also, we show that the orbital
eccentricity of the outermost planet is actually ill-determined, although bounded by $\sim
0.2$. Finally, we investigate possible orbital non-coplanarity of the system, and limit
the mutual inclination between the planets \emph{b} and \emph{c} orbits by
$5^\circ-15^\circ$, depending on the angular position of the mutual orbital nodes.
\end{abstract}

\section{Introduction}
\label{intro}
The first extrasolar planet in the system orbiting the red dwarf GJ876 was detected
independently by \citet{Delfosse98} and \citet{Marcy98} on the basis of high-precision
radial velocity (hereafter RV) Doppler measurements, acquired at ELODIE+CORALIE and Keck
HIRES spectrographs. Soon after this, the second planetary companion was discovered by
\citet{Marcy01}. That discoveries showed that the exoplanetary system of GJ876 is an
extraordinary object. The two planets \emph{b} and \emph{c} were massive giants (having
minimum masses $m\sin i$ of roughly $2$ and $0.6$ of Jupiter masses) orbiting in short
period orbits (approximately $60$ and $30$ days). Therefore, it became clear that the
planets orbital dynamics should be significantly affected by the 2/1 mean-motion resonance
(hereafter MMR). The resonance itself is not yet very astonishing, because now we already
know many extrasolar planets in various MMRs. The extraordinarity of the GJ876 system
comes from the fact that interplanetary gravitational perturbations were directly detected
in the observed radial velocity data, i.e. they reveal themselves on the observational
time scale \citep{Laughlin01,Rivera01}.

Still, no other extrasolar planetary system around a main sequence star is known to
demonstrate so clear and well-measurable signatures of planetary perturbations in its
radial velocity time series (although, there are some exoplanetary systems where such
perturbations are suspected to be non-negligible). A few favoring factors meet each other
in GJ876: large masses of the planets \emph{b} and \emph{c}, small star mass (hence
especially large planet/star mass ratios), rather short orbital periods (thus shorter
perturbations time scale), and, of course, the low-order MMR. The dynamical perturbations
reveal themselves mainly in the form of secular circulation of the planetary apsidal
lines, triggering slow change in the non-sinusoidal shape of the RV oscillations being
observed. The notion ``secular'', however, looks somewhat odd here, since the period of
this ``secular'' circulation is only about $10$~yrs. During the whole observation term
since 1998 till today, these apsidal lines completed roughly a single revolution each.

This dynamical effect complicates the procedure of the RV data analysis, but in exchange
it enables us to determine (solely from the RV data) the inclination of the system to the
sky tangent plane. This allows us to determine, instead of the minimum planetary masses
$m\sin i$, the true masses $m$, which normally remain unconstrained in the RV exoplanet
detections.

Later, \citet{Rivera05} reported the discovery of the third planet in the system, based on
further Keck RV observations of GJ876. This planet (\emph{d}) possesses very low mass of
$\sim 7.5$ Earth masses and a very short orbital period of approximately $2$~days. In the
paper \citep{Bean09}, the question of orbital non-coplanarity of this system was studied
extensively, and these authors gave an estimation of $\sim 5^\circ$ for the mutual orbital
inclination of the planets \emph{b} and \emph{c}. This result was based on the Keck RV
data from \citep{Rivera05} and on the HST astrometry data from \citep{Benedict02}.
Recently, \citet{Correia10} reanalyzed these old Keck data adding to them a new array of
very accurate RV measurements, obtained by the HARPS spectrograph (ESO). He presented
improved three-planet orbital fits, which generally agree with the fits by
\citet{Rivera05}.

Finally, in the very recent paper \citep{Rivera10} the discovery of the fourth Uranus-mass
planet \emph{e} was reported. This last planet is also remarkable, because it forms a
Laplace three-planet resonance with two other giant planets, so that the ratio of the
periods $P_c:P_b:P_e$ is close to $1:2:4$.

It is interesting, however, that the RV signature of this fourth planet had been found in
the old Keck data already, e.g. in \citep{Baluev08-IAUS249}. It is notable that the fourth
planet parameters from \citep{Baluev08-IAUS249} are surprisingly close to the ones given
by \citet{Rivera10}, including even the mean longitude and the small orbital eccentricity.
This orbital configuration appears pretty stable at long time scales. \citet{Rivera10}
also mention that they were seeing the signs of the fourth planet in their data since
2005, but until 2010 they could not find a stable four-planet solution, due to the large
planet \emph{e} eccentricity. The reason of such difference probably comes from the fact
that \citet{Rivera10} did not take into account the annual errors of a few m/s, which were
detected in the old Keck data by \citet{Baluev08-IAUS249}. Such annual errors may appear
relatively frequently in the RV planet search surveys, as discussed in \citep{Baluev09a}.
They may have different physical sources; in case of GJ876 they were probably caused by
some errors in the data reduction pipeline, since new Keck data from \citet{Rivera10} look
free from such errors. In fact, a careful error analysis of the RV data could enable the
robust detection of the fourth planet several years ago already.

Since the quality of the Keck data is considerably improved in \citep{Rivera10}, and the
completely new very accurate RV data is now published \citep{Correia10}, we conclude that
it is time to perform such detailed analysis now. Accurate and unbiased values of
planetary masses and orbital parameters in so remarkable system as GJ876 represent a huge
interest for the celestial mechanics studies, as well as for more astrophysical branches
like the tidal star-planet interaction studies (note the ``hot super-earth'' GJ876
\emph{d}) or for the planet formation theories. In particular, it is important to have
some reliable estimations or limits of the orbital non-coplanarity of such system, or to
know how large the orbital eccentricity of the ``hot Earth'' GJ876 \emph{d} actually can
be.

The plan of the paper is as follows. First, in Sect.~\ref{sec_fit} we describe in detail
the fast approach of fitting the dynamically perturbed planetary configurations on the
basis of the star RV time series. We also explain the conventions we adopt to determine
the reference osculating orbital parameters, and describe the numerical integrator we use
in this study. In Sect.~\ref{sec_data} we describe in more details the RV data we use in
the paper and also give some preliminary analysis results, obtained during direct fitting
of these data. In Sect.~\ref{sec_redj}, we find that these RV data cannot be modeled in a
traditional way, because they contain significant fraction of autocorrelated (non-white)
noise. Also, we propose here a modified RV fitting algorithm, which allows to take such
correlated RV errors into account properly. In Sect.~\ref{sec_fourth} we study the planet
\emph{e} orbital parameters and show that its eccentricity is still ill-determined
although bounded by $0.2$ from the upper side. In Sect.~\ref{sec_nom} we give final
estimations of all planetary parameters, taking into account the effects of the RV noise
correlateness and of the bad determinability of the planet \emph{e} eccentricity. In
Sect.~\ref{sec_inc}, we study the (non-)coplanarity of the system in view of currently
available RV data. Finally, Sect.~\ref{sec_dyn} describes briefly the dynamical evolution
of the admissible GJ876 orbital configurations.


\section{Fitting graviationally perturbed orbits}
\label{sec_fit}
\subsection{Evaluating radial velocity function}
\label{sec_fit_RV}
Before we proceed to the investigation of the GJ876 planetary system itself, we describe
the algorithm we used to perform $N$-body fitting of RV data. Although such $N$-body
fitting have been already carried out in several previous works
\citep{Laughlin01,Rivera01}, but they omit important details of the algorithms used. We
may highlihgt a very recent paper \citep{Pal10}, which introduces a method of orbital
fitting, which is similar (in spirit) to the method that we use here. However, many ideas
of the methods developed by \citet{Pal10} look pretty different, so we still need to
describe our method in detail here. The algorithm that we use represents a variation of
the popular non-linear Levenberg-Marquardt $\chi^2$ minimization algorithm with a minor
modification necessary to take into account the ``RV jitter'' phenomenon, as discussed in
\citep{Baluev09a} and below. This algorithm is an iterative gradient method, which
requires to evaluate (on different iteration stages) the fittable model (RV curve model,
in our case) and its partial derivatives over all free parameters. In case of GJ876, the
two latter subtasks are non-trivial, since we need to take into account the full dynamical
model of planetary perturbations.

Let us write down the equations of the $N$-body problem in the astrocentric coordinate
system:
\begin{equation}
\frac{1}{k^2}\frac{d^2\vec r_i}{dt^2} = - (1+\mu_i) \frac{\vec r_i}{r_i^3} +
\sum_{j=1..\mathcal N\atop j\neq i} \mu_j \left(
\frac{\vec r_j - \vec r_i}{|\vec r_j - \vec r_i|^3} - \frac{\vec r_j}{r_j^3}
\right), \quad i = 1,2,\ldots,\mathcal N.
\label{moteq}
\end{equation}
Here $\mu_i$ are the ratios of the planet masses $m_i$ to the star mass $M_\star$, and
$k^2=GM_\star$ is the analogue of the Gauss gravitational constant. Note that further we
may call $\mu_i$ just ``planetary masses'' for shortness, since usually it should not
introduce any misunderstandings. We assume that $z$ axis is directed along the observer's
line of sight, and two other axes are directed arbitrarily in the tangent sky
plane.\footnote{Note that the term ``$N$-body'', which we use in the paper frequently,
does not imply that the number of the \emph{bodies} in the system is denoted by $N$.
Instead, $\mathcal N$ stands for the number of interacting \emph{planets}, $N$ for the
number of RV measurements, and ``$N$-body'' must be understood as a solid term.}

The initial conditions for the system~(\ref{moteq}) may be expressed via some set of
$6\mathcal N$ osculating orbital elements forming $\mathcal N$ vectors $\vec\pi_i$ using
the well-known functions of the Keplerian motion. However, there is no unique definition
of the osculating Keplerian elements. Their set depends not only on our choice of the
parametrization, but also on the non-unique splitting of the forces in~(\ref{moteq}) into
the unperturbed Keplerian and perturbational parts. For instance, we can assume that the
Keplerian part of the force is $-k^2\vec r_i/r_i^3$, and everything else in the right hand
side of~(\ref{moteq}) is the perturbation. In such a case, the initial planetary
coordinates and velocities could be expressed via corresponding osculating orbital
elements as
\begin{equation}
\left. \vec r_i \right|_{t=0} = k^{\frac{2}{3}} \vec K_r(\vec\pi_i), \qquad
\left. \vec v_i \right|_{t=0} = k^{\frac{2}{3}} \vec K_v(\vec\pi_i),
\label{Kep}
\end{equation}
where the functions $\vec K_r$ and $\vec K_v$ should represent the solution of the
standard equation of the Keplerian motion $\ddot{\vec r}=-\vec r/r^3$ at $t=0$. Each
vector $\pi$ incorporates six osculating orbital elements of the planets: orbital period
$P$, eccentricity $e$, pericenter argument $\omega$, mean argument of latitude $\psi$
(mean anomaly + $\omega$), inclination to the sky plane $i$, longitude of the ascending
node $\Omega$. The specific of the exoplanetary orbit determination problem sets a few
pecularities of these parameters that we need to highlight:
\begin{enumerate}
\item We remind the reader that the angle $\omega$ is traditionally counted from the
\emph{descending} orbital node \citep[][sect.~2.2]{FerrazMello-lec1}. Historically, this
deviation from the classical definition of $\omega$ occured because exoplanet researchers
traditionally model the host \emph{star}'s RV wobble using the Keplerian formula $V = K
(\cos(\omega+\upsilon)+e\cos\omega)$, which actually expresses the RV of the orbiting
\emph{planet} itself. This shifts the derived value of $\omega$ by $\pi$, due to the RV
sign change. Alternatively, we may think that $\omega$ refers to the \emph{star}'s orbit
around the common barycenter, although such treatment becomes a bit vague for multi-planet
systems. This pecularity of the angle $\omega$ is insignificant for the most of the
exoplanets, since the positions of the planetary orbital nodes usually remain
unconstrainable anyway. It may become important for GJ876, however.
\item Consequently, the angle $\psi$ is also counted from the descending node. We use
$\psi$ instead of the full mean longitude $\lambda=\Omega+\psi+\pi$, since the absolute
positions of the orbital nodes still remain unconstrained until the astrometric
observations are used. Radial velocities can only constrain the differences between
$\Omega_i$. For coplanar configurations, dealing with $\psi$ and $\omega$ is equivalent to
dealing with $\lambda$ and the pericenter longitude $\varpi=\Omega+\omega+\pi$, since all
$\Omega_i$ are equal to the same constant value.
\item In the equations~(\ref{Kep}), the dependence on the actual $k$ (in fact, on
$M_\star$) is extracted separately in the external factors $k^{2/3}$. Such dependence on
$k$ appears when $\vec\pi$ contains planetary orbital period $P$ (or e.g. mean motion) as
a primary parameter, and the semi-major axis is treated as only a derived one. We use such
``period-oriented'' choice because it is the planetary orbital period (or the period of
corresponding RV oscillation) which is drawn from the observations directly, rather than
the orbital semi-major axis. If we chose the semi-major axis as a primary parameter, the
factors in~(\ref{Kep}) would be $1$ and $k$.
\end{enumerate}

The astrocentric coordinate system makes the $N$-body equations~(\ref{moteq}) simple for
numerical integration, but it is not well suitable to reference orbital elements. This is
mainly because of relatively significant dependence of the planetary orbital periods on
the choice of the coordinate system, which has been already discussed in
\citep{LissRiv01,LeePeale03,Baluev08c}. As it was noted in these works, the osculating
orbital periods (or mean motions) referenced in the Jacobi coordinate system generally
match better the apparent orbital periods (or apparent average drift of the mean
longitude). In practice that would mean that, e.g., we may freely experiment with turning
on/off the gravitational interactions of any planet without the need to manually adjust
its orbital period when switching from one model to another. Otherwise, we should remember
that there may be a significant offset between the best fitting orbital period in the
Keplerian RV model framework (which implies apparent orbital period) and the full $N$-body
framework (the osculating period). Although such offsets are small ($\mathcal O(\mu_i)$),
they may be comparable to or even significantly exceed the relative statistical error of
the respective periods (e.g. $\sim 10^{-5}$ in case of the innermost planet of GJ876). In
such a case, the orbital fitting is slowed down and eventually may even fail to converge
to the correct period value (it may be attracted by a spurious close alias or just noisy
neighboring period). The choice of the Jacobi system usually eliminates these troubles, if
we also assume that each osculating Keplerian orbit refers to a fictitious central mass
incorporating the star's and this planet's mass and masses of all inferior planets.

According to \citep[appendix A]{Baluev08c}, one way to define the unperturbed part of the
$N$-body Hamiltonian in the Jacobi coordinates is
\begin{equation}
H_{\mathrm{Kep},i} = \frac{\gamma_i}{\gamma_{i-1}} \frac{{\vec p'_i}^2}{2\mu_i} -
  \gamma_{i-1} \frac{k^2\mu_i}{r'_i},
\quad \gamma_i=1+\sum_{j=1}^i \mu_i.
\label{HamKep}
\end{equation}
We adopt this (non-unique) way to split the Hamiltonian into the Keplerian and
perturbational parts, because the corresponding equation of the Keplerian motion
\begin{equation}
\frac{d^2\vec r'_i}{dt^2} = -{k'_i}^2 \frac{\vec r'_i}{{r'_i}^3}, \qquad
{k'_i}^2 = k^2\gamma_i = GM_\star\gamma_i
\end{equation}
involves the value of the central mass $M_\star \gamma_i$ instead of $M_\star$. This is
exactly what we need. The initial conditions for the Jacobi coordinates and velocities
should represent a solution of the latter equation and therefore should look like
\begin{equation}
\left. \vec r_i' \right|_{t=0} = {k'_i}^{\frac{2}{3}} \vec K_r(\vec\pi_i), \qquad
\left. \vec v_i' \right|_{t=0} = {k'_i}^{\frac{2}{3}} \vec K_v(\vec\pi_i)
\label{KepJ}
\end{equation}
These formulae differ from~(\ref{Kep}) only by the coefficient $k$, which now involves
planetary masses too. Note that in case of a highly hierarchical system with negligible
mutual perturbations between planets, such definition of the osculating orbital periods
would make them infinitesimally close to the apparent revolution periods. The exoplanetary
systems are not always hierarchical, but in practice (in particular, for GJ876) the
offsets between apparent and so-defined osculating planet periods in $\vec\pi_i$ usually
remain satisfactory.

After that, we can transform the Jacobi vectors~(\ref{KepJ}) to the astrocentric ones
using the formulae
\begin{eqnarray}
\vec r_i &=& \vec T_i(\vec r'_1,\ldots,\vec r'_i, \mu_1,\ldots,\mu_{i-1}) \equiv
 \vec r'_i + \sum_{j=1}^{i-1} \frac{\mu_j}{\gamma_j} \vec r'_j, \nonumber\\
\vec v_i &=& \vec T_i(\vec v'_1,\ldots,\vec v'_i, \mu_1,\ldots,\mu_{i-1}).
\label{JtoH}
\end{eqnarray}
The formulae~(\ref{KepJ}) and~(\ref{JtoH}) jointly express the Cartesian initial
conditions in the astrocentric system via the osculating orbital elements $\vec\pi_i$
defined in the Jacobi system, and via the planet masses $\mu_i$. For shortness, we can
write down these compound dependences as
\begin{eqnarray}
\left. \vec r_i \right|_{t=0} &=& \vec R_i(k,\vec\pi_1,\ldots,\vec\pi_i, \mu_1,\ldots,\mu_i), \nonumber\\
\left. \vec v_i \right|_{t=0} &=& \vec V_i(k,\vec\pi_1,\ldots,\vec\pi_i, \mu_1,\ldots,\mu_i).
\label{initcond}
\end{eqnarray}
Note that these functions also involve the parameter $k$, which depends on the star mass
$M_\star$, which we assume is \emph{a priori} known from astrophysical models of its
spectrum. Following \citet{Correia10}, we assume $M_\star=0.334M_\odot$ throughout the
paper.

Having the astrocentric initial conditions from~(\ref{initcond}), we can
integrate~(\ref{moteq}) to the desired time. After that, we will have astrocentric
planetary positions and velocities. We prefer to integrate the astrocentric motion
equations~(\ref{moteq}), since the analogous equations in the Jacobi system would be
significantly more complicated and thus would slow the algorithm down. After the
integration, the barycentric velocity vector of the star can be obviously expressed via
the planetary astrocentric velocities as
\begin{equation}
\vec v_\star = - \left. \sum_{i=1}^{\mathcal N} \mu_i \vec v_i \right/ \left(1+\sum_{i=1}^{\mathcal N} \mu_i \right).
\end{equation}
Finally, all stages explained in this section allow us to determine the model RV curve,
based on the planetary orbital parameters $\vec\pi_i$ and masses $\mu_i$ as fittable free
variables. The total observable radial velocity of the star incorporates also a constant
radial velocity (we denote it as $c_0$), caused by the motion of the planetary system
barycenter. Moreover, since the RV measurements available for GJ876 are basically
relative, we should also introduce different fittable values of $c_0$ for the data series
obtained at different instruments.

\subsection{Using variational equations}
\label{sec_fit_var}
Probably the most important stage of the Levenberg-Marquardt algorithm (as well as of any
other gradient optimization method) is evaluation of the partial derivatives of the data
model over the free parameters. In problems with complicated model function (like $N$-body
RV fitting) most of the computational time is spent during evaluation of these
derivatives. The simple way to evaluate them is just to use numerical differentiation of
the original RV model (which should be calculated using $N$-body integration). This way is
easy to implement algorithmically, but this leads to a very bad speed/error ratio, since
the error of numerical differentiation is considerably larger than the error of the
original function. A better way to evaluate these derivatives is to integrate variational
equations of the $N$-body problem (in the literature on non-linear optimization they are
also called sensitivity equations, see e.g. chapter~8 in the book by \citet{Bard}). It
allows to obtain roughly the same accuracy of derivatives as in the original function with
the same integration step and number of equations to integrate. Planetary positional
vectors $\vec r_i$ and their velocities $\vec v_i=d\vec r_i/dt$ depend on the time, on the
planetary masses $\mu_i$ and on the orbital parameters $\vec\pi_i$. Differentiation
of~(\ref{moteq}) over $\vec\pi_i$ and $\mu_i$ yields the following linear ODE system for
partial derivatives of $\vec r_i$:
\begin{eqnarray}
\frac{1}{k^2} \frac{d^2}{dt^2}\frac{\partial\vec r_i}{\partial\vec\pi_j} &=&
 - (1+\mu_i)\, \tens\Pi(\vec r_i) \frac{\partial\vec r_i}{\partial\vec\pi_j} +\nonumber\\
& & + \sum_{k=1..\mathcal N\atop k\neq i} \mu_j\left[
 \tens\Pi(\vec r_j-\vec r_i)\left( \frac{\partial\vec r_j}{\partial\vec\pi_j} -
 \frac{\partial\vec r_i}{\partial\vec\pi_j} \right) -
 \tens\Pi(\vec r_j)\frac{\partial\vec r_j}{\partial\vec\pi_j} \right], \nonumber\\
\frac{1}{k^2} \frac{d^2}{dt^2}\frac{\partial\vec r_i}{\partial \mu_j} &=&
- \frac{\vec r_j}{r_j^3} + (1-\delta_{ij})
\frac{\vec r_j-\vec r_i}{|\vec r_j-\vec r_i|^3} -
 (1+\mu_i)\, \tens\Pi(\vec r_i) \frac{\partial\vec r_i}{\partial \mu_j} + \nonumber\\
& & + \sum_{k=1..\mathcal N\atop k\neq i} \mu_j\left[
 \tens\Pi(\vec r_j-\vec r_i)\left( \frac{\partial\vec r_j}{\partial \mu_j} -
 \frac{\partial\vec r_i}{\partial \mu_j} \right) -
 \tens\Pi(\vec r_j)\frac{\partial\vec r_j}{\partial \mu_j} \right], \nonumber\\
& & \tens\Pi(\vec a) = \frac{a^2\Id - 3 \vec a\otimes\vec a}{a^5}, \qquad
\delta_{ij} = \left\{ {1,\, i=j\atop 0,\, i\neq j}\right. .
\label{senseq}
\end{eqnarray}
Here $\delta_{ij}$ is the Kronecker delta, $\tens I$ is the identity matrix, $\otimes$
denotes the dyadic product of vectors, and thus $\tens\Pi(\vec a)$ is a $3\times 3$ matrix
having elements $(\delta_{ij}-3a_i a_j/a^2)/a^3$. The derivatives $\partial\vec
r_i/\partial\vec\pi_j$ should be treated as $3 \times 6$ Jacobian matrices. The
system~(\ref{senseq}) should be integrated simultaneously with the original
equations~(\ref{moteq}). The partial derivatives of $\vec v_i$ are equal to the temporal
derivatives of the corresponding partial derivatives of $\vec r_i$ and can be obtained
during the integration automatically. The initial values for $\partial\vec
r_i/\partial\vec\pi_j$ and $\partial\vec v_i/\partial\vec\pi_j$ can be obtained by means
of differentiation of the compound functions~(\ref{initcond}) over the corresponding
parameters. We do not give here the resulting expressions since they are rather
complicated and numerous, and actually their main parts involving derivatives of $\vec
K_r, \vec K_v$ over $\vec\pi_i$ can be found in the classical literature on orbit
refinement \citep[e.g.][sect.~3.3]{Duboshin76}. The derivatives of~(\ref{initcond}) over
$\mu_i$ are also clearly calculatable, but they are rather complicated as well, and we
have to omit them too.

\subsection{Numerical integrator}
\label{sec_fit_int}
We need some numerical integration method to solve the differential systems~(\ref{moteq})
and~(\ref{senseq}). We do not need it to be well-behaved (in any sense) on long time
scales, since our integration timescale is rather short ($\sim 10$~yrs or $\sim
100$~revolutions of the inner giant planet). But we do need it to be fast on this
timescale. After a few experiments, we suggest that the integrators of the Everhart type
\citep{Everhart73,Everhart74} would meet our requirements. We say ``Everhart type'' since
we actually used several integrators belonging to a general family of integrators similar
to those constructed originally by Everhart. The original Everhart integrator was based on
the Radau quadrature formula, based on certain asymmetric splitting of the integration
step. In accordance with \citet{Avdushev10}\footnote{See preprint at
\emph{http://www.scharmn.narod.ru/AVD/Gauss\_15\_2.pdf}, in Russian}, it is possible to
construct an integrator based on an arbitrary splitting of the integration step. These
integrators represent, in fact, implicit Runge-Kutta integrators, equipped by an efficient
method of evaluating the predictors on each step. Certain segment splittings (like the
Radau one) allow to increase the order of the integrator significantly. Moreover, it is
known that the use of the Legendre splitting makes the integrator symplectic. Symplectic
integrators are very well suitable for long-term integrations, because they can preserve
the Hamiltonian properties of the original $N$-body problem over a much longer term (e.g.,
they show much slower energy error accumulation). It is important that in the case of the
Everhart integrators the symplecticity can be obtained even with no significant trade-off
or undesired side effects.

The only significant disadvantage of such integrators is that they require constant time
step for symplecticity. The algorithm itself does contain an optional step-size control
mechanism, but variable step destroys the symplectic property. Therefore, we choose to use
the same symplectic integrator both for the short and long integration terms. This is the
$16$-th order Everhart type integrator, based on the $8$-node Legendre splitting. For
short-term integrations (needed to perform orbital fits) we use the variable step. For
long-term integrations (stability tests), we hold the step size fixed.

The algorithm that we use is largely based on the Avdyushev's FORTRAN program
code\footnote{See \emph{http://www.scharmn.narod.ru/AVD/Gauss\_15.for} and\\
\emph{http://www.scharmn.narod.ru/AVD/GAUSS\_32.for}} with several modifications, aimed to
increase its performance. We omit the detailed description here, since we plan to provide
it in a separate work in the future, along with the program source code.

\section{Radial velocity data for GJ876 and its preliminary planetary orbital solution}
\label{sec_data}
The main datasets available are $162$ Keck RV measurements from \citep{Rivera10}, and $52$
HARPS RV measurements from \citep{Correia10}. \citet{Correia10} also use RV datasets
obtained at ELODIE and CORALIE spectrographs, and there is also Lick RV dataset in
\citep{Marcy01}. We include these data too, although it is necessary to note that they are
considerably less accurate and in fact have insignificant effect on the system orbital
solution~--- only a few per cent of the parameters uncertainties. The average internal RV
precision of the HARPS and Keck datasets is about $1$ and $2$~m/s, their time span about
$4.7$ and $12.6$ years, respectively. The average RV uncertainties of the Lick, ELODIE,
and CORALIE datasets are about $10$~m/s or more, and they span $6$ and $8$ years and $3$
months, respectively.

The internal errors stated in the published RV tables do not yet constitute the full RV
uncertainty. It is well-known that high-precision RV exoplanet searches suffer from the
phenomenon of the so-called RV ``jitter'', which adds extra irregular variations in their
RV data. This excessive jitter was initially explained via star astrophysical activity
effects, but later it was shown that various systematic instrumental effects
\citep{Baluev09a} may be significant as well (and may even dominate sometimes). Anyway,
this jitter has to be taken into account during any further analysis. It is important here
that in practice the effective value of the apparent RV jitter of the same star is usually
different for different instruments, implying different (and poorly known) weighting
coefficients for different datasets. To take this jitter into account properly, we utilize
the maximum-likelihood approach suggested in \citep{Baluev09a}. This approach is based on
maximizing the datasets' joint likelihood function, which depends on the RV curve
(planetary) parameters as well as on the parameters determining the statistical structure
of the errors in the RV data. This allows to estimate all necessary jitter values
simultaneously with planetary parameters.

According to \citep{Baluev08c}, we will obtain the best fitting values of all involved
parameters by means of maximizing the objective function
\begin{equation}
\ln \tilde{\mathcal L} = - \sum_{j=1}^J \sum_{i=1}^{N_j} \left(
 \ln\sigma_{\mathrm{full},ji} + \frac{1}{2\gamma} \left(\frac{r_{ji}}{\sigma_{\mathrm{full},ji}}\right)^2
 \right) - N\ln\sqrt{2\pi},
\label{lik}
\end{equation}
where $N_j$ stands for the number of the data points in the $j$th dataset, $J$ denotes the
number of the datasets ($J=5$ in our case), $N=\sum N_j$ is the total number of
observations, $r_{ji}$ is the $i$th RV residual (O-C) in the $j$th dataset, and
$\sigma_{\mathrm{full},ji}^2 = \sigma_{\star,j}^2 + \sigma_{\mathrm{meas},ji}^2$ is the
full variance of the corresponding RV measurement, which incorporates the internal
measurement variance $\sigma_{\mathrm{meas},ji}^2$ as well as the jitter variance
$\sigma_{\star,j}$. The quantity $\gamma$ (not to be mixed with gammas in~(\ref{HamKep}),
which are indexed) is equal to $1-d/N$, where $d$ is the number of free parameters of the
RV curve (number of degrees of freedom in the RV model). This correcting divisor helps to
remove the systematic bias in the RV jitter estimations, as discussed in
\citep{Baluev09a}. This is necessary, because the residuals $r_{ji}$ to the best fitting
model are always systematically smaller than the original data errors, and without any
correction they would yield underestimated value for the jitter.

The function~(\ref{lik}) depends on the RV curve parameters (via $r_{ji}$, which involve
the RV model) and on the RV jitter values (via $\sigma_{\mathrm{full},ji}$). The values of
these parameters, where the function~(\ref{lik}) reaches its maximum, represent the
necessary best fitting estimations. To estimate a quality of a given fit, we use the
statistic $\tilde l$ defined in \citep{Baluev09a}. This quantity represents a monotonous
function of $\tilde{\mathcal L}$, but is measured in m/s and thus is intuitively more
clear and comparable to more traditional measures like r.m.s. (which we will use too, for
some comparison).

\begin{figure}
\center
 \includegraphics[width=0.75\textwidth]{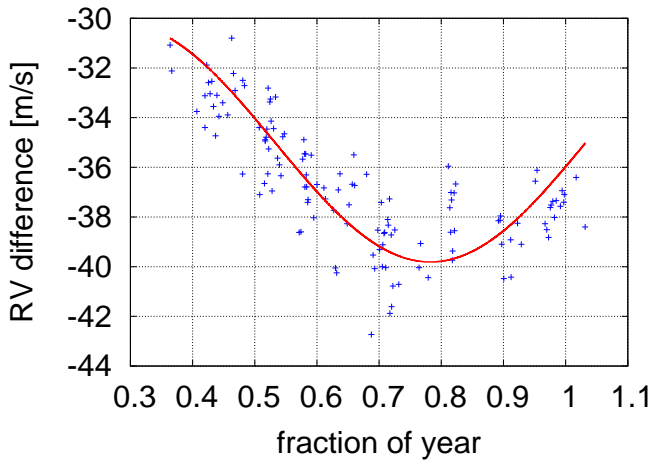}
\caption{The points mark the differences between the old \citep{Rivera05} and new
\citep{Rivera10} Keck RV measurements for GJ876, phased to the one year period (unit means
end of a year). These differences show clear systematic variation of about $8$~m/s in
total, which is well above the typical measurement uncertainty of $3-4$~m/s (old data) and
$1-2$~m/s (new data). This variation was probably caused by errors in the data reduction
pipeline used by \citet{Rivera05}. The solid curve represents the graph of the annual
systematic variation, which was estimated in \citep{Baluev08-IAUS249} on the basis of the
old Keck data.}
\label{fig_Keck_diff}
\end{figure}

Other details of this method can be found in \citep{Baluev08c,Baluev09a}. Here we use the
same formalism, with the major difference that we now deal with the full gravitational
$N$-body RV model, instead of the multi-Keplerian one. Also, on contrary with
\citep{Baluev08c}, we do not add in the RV model the terms describing possible annual
systematic errors in the data. Although such annual errors existed in the old Keck data
from \citep{Rivera05}, it seems the new data should have been fully corrected (see
Fig.~\ref{fig_Keck_diff}). The HARPS observations for other stars seemingly were always
free from such errors.

For each planet, we will fit the following osculating orbital parameters: orbital period
$P$ (equiv. to the mean motion), mean argument of latitude $\psi$, eccentricity $e$,
pericenter argument $\omega$. We also fit common orbital inclination $i$ (assuming
coplanar system). In addition to the orbital parameters, we should also fit the planet
masses. However, it is not convenient to adopt planet masses as primary fit parameters. In
the traditional non-perturbed framework, the masses themselves are not determinable at
all. In such a case, the RV oscillation semi-amplitude, $K$, is fitted (as primary
parameter), which allows to derive the minimum planet mass $m\sin i$. It is well-known
that without detectable interplanetary gravitational perturbations the inclination $i$
remains unconstrained, and the true planet mass $m$ remains unknown. In our case the
system inclination is well-constrainable, but only for coplanar model. We will consider
non-coplanar configurations below too, where individual inclinations may be poorly
determined. In such a case it would be better not to mix the uncertainties of the RV
amplitude and inclination, otherwise we risk to deal with significant troubles during
numerical fitting, due to strong correlations between various fit parameters. Also, we
usually will not include the innermost planet in the $N$-body integration, assuming that
on the observation timescale its motion is close to a Keplerian orbit. For this planet, we
just have to leave with the RV semi-amplitude.

Therefore, it would be better still to adopt the RV semi-amplitudes as primary fit
parameters, instead of the planet masses. But then we must clarify what is ``RV
semi-amplitude'' in the perturbed case. Actually, we cannot strictly define any
``amplitude'' for the perturbed motion, since this motion is not periodic. Nevertheless,
in the case of GJ876 the deviations from the strict periodicity are small, and we still
can determine the RV amplitude approximately, within the error of $\mathcal O(\mu_i)$.
Eventually, we need just to bind this RV amplitude parameter to the corresponding planet
mass. Therefore, we can just define it via $m_i$ using some simple formula close in shape
to the formula from the Keplerian RV case. We adopt the following definition:
\begin{equation}
\mu = \frac{K \sqrt{1-e^2}}{\sin i} \left(\frac{P}{2\pi k^2}\right)^{1/3}.
\label{mass_def}
\end{equation}
We must emphasize that this formula no longer \emph{expresses} planet mass via the
corresponding RV semi-amplitude, as it would be in the case of unperturbed motion ($\mu_i
\ll 1$). Rather, it now \emph{defines} the ``RV semi-amplitude'' via the planet mass. We
may think of~(\ref{mass_def}) as of a definition of the \emph{osculating} RV
semi-amplitude, which completes the set of usual osculating orbital elements. So-defined
``osculating RV semi-amplitude'' is in fact just an intermediate fit parameter needed to
separate the uncertainty of the orbital inclination from the uncertainty of the planet
mass. The apparent semi-amplitude of the star's RV oscillation, caused by the
corresponding planet, should be very close to the value of $K$ defined
in~(\ref{mass_def}). Furthermore, we decide to use not even the semi-amplitude $K$ itself,
but the quantity $\tilde K = K\sqrt{1-e^2}$, as it was done in \citep{Baluev08c}. Such
choice allows to eliminate the eccentricity from the relation~(\ref{mass_def}) and thus
facilitates the conversions between $\mu$ and $\tilde K$.

\begin{table}
\caption{Best fitting coplanar orbital solution for GJ876 system (epoch JD2452000).}
\begin{center}
\begin{tabular}{@{}llllll@{}}
\hline\noalign{\smallskip}
parameter            & planet b (*)    & planet c (*)    & planet d        & planet e (*)    & \\
\multicolumn{5}{c}{fitted planetary parameters} & \\
$P$~[days]           & $60.9904(68)$   & $30.1829(63)$   & $1.937886(18)$  & $124.51(52)$    & \\
$\tilde K$~[m/s]     & $213.21(34)$    & $84.65(36)$     & $6.18(29)$      & $3.41(33)$      & \\
$\psi$~[$^\circ$]    & $341.13(20)$    & $71.09(46)$     & $357.6(3.2)$    & $299.3(7.3)$    & \\
$e$                  & $0.0328(13)$    & $0.2498(28)$    & $0.178(44)$     & $0.008(27)$     & \\
$\omega$~[$^\circ$]  & $248.7(2.9)$    & $252.08(51)$    & $224(16)$       & $181(77)$       & \\
$i$~[$^\circ$]       & \multicolumn{4}{c}{$56.1(1.5)$} & \\
\multicolumn{5}{c}{derived planetary parameters} & \\
$m$~[$M_{Jup}$]      & $2.377(42)$     & $0.747(13)$     & $0.0218(11)$    & $0.0482(47)$    & \\
$a$~[AU]             & $0.211018(16)$  & $0.131727(18)$  & $0.02110625(13)$& $0.33961(94)$   & \\
\noalign{\smallskip}\hline\noalign{\smallskip}
\multicolumn{6}{c}{RV data series and general fit parameters}\\
                     & Keck            & Lick            & HARPS           & ELODIE          & CORALIE         \\
$c_0$~[m/s]          & $50.95(27)$     & $-31.4(4.9)$    & $-1337.87(42)$  & $-1864.1(3.7)$  & $-1904.0(4.6)$  \\
$\sigma_\star$~[m/s] & $2.37(22)$      & $-11.3(6.0)$    & $1.63(23)$      & $21.3(3.2)$     & $19.7(4.3)$     \\
r.m.s.~[m/s]         & $3.00$          & $27.6$          & $1.84$          & $33.5$          & $32.0$          \\
\multicolumn{5}{c}{$\tilde l=5.842$/$2.777$~m/s, $d=26$}\\
\noalign{\smallskip}\hline
\end{tabular}
\end{center}
\smallskip
Each estimation is accompanied by its uncertainty in parenthesis (e.g., $0.30(10)$ means
$0.30\pm 0.10$, and $30.0(1.0)$ means $30.0\pm 1.0$). These uncertainties were calculated
from the Fisher matrix of the likelihood function \citep{Baluev09a}. The uncertainties of
the planet masses and semi-major axes do not incorporate stellar mass uncertainty. Planets
included in the $N$-body integration are marked with (*). The negative value for the Lick
RV jitter means (symbolically) that the corresponding value of $\sigma_\star^2$ is
actually negative. The second value of $\tilde l$ refers to the same fit based on only
Keck and HARPS data (which offers practically the same estimations within a few per cent
of the uncertainties).
\label{tab_prelim}
\end{table}

To obtain some basic preliminary estimations of the system parameters, we take the
Keck-only orbital solution from \citep{Rivera10} as a starting approximation and perform
the non-linear maximization of the likelihood function, as it was explained above.
Table~\ref{tab_prelim} contains the resulting best fitting estimations of all planetary
parameters and RV jitter for different datasets. Note that here we exclude the innermost
planet \emph{d} from the integration, assuming that it moves along a Keplerian orbit in
the common system orbital plane. Its orbital period is considerably smaller than periods
of other planets, as well as its mass, and therefore it does not show significant
dynamical interaction with other planets on the observational timescale. Taking this
planet into $N$-body integration slows down the calculations dramatically. To ensure the
gravitational influence of this planet is insignificant, we performed a similar (much
longer) calculation, based on the full four-planet $N$-body model. Almost all of the
resulting best fitting parameters were practically identical, with negligible offsets of
no more than $3\%$ of the corresponding uncertainties. The only exception was the period
of this same planet \emph{d}, which decreased by $1.9\cdot 10^{-5}$~day, i.e. roughly by
its uncertainty. From the statistical view point, such shift should not be neglected,
since it would correspond to rather large one-sigma significance
level.\footnote{Throughout this paper, we will usually use the popular ``$n$-sigma'' style
to specify various confidence probabilities. For example, we will say that a given
statement is valid at an $n$-$\sigma$ significance level, when corresponding confidence
probability is equal to the probability for a Gaussian random variable to deviate from its
mean by no more than $n$ times its standard deviation (``sigma''). The significance levels
of $1,2,3$-$\sigma$ correspond to the confidence probabilities of, respectively, $68.3\%$,
$95.4\%$, and $99.73\%$ (closer to $100\%$ means more significant).} This shift is
nevertheless extremely small. It does not affect the motion of the other planets at a
measurable level. Since the dynamical effects due to the innermost planet are so small, we
neglect them below, and integrate only the system of three remaining planets.

The apparent orbital period $P_d$ is also shifted due to the planetary aberration (also
known as Roemer effect), which is caused by the finite light speed and is similar (in
origin) to the Doppler effect \citep{FerrazMello-lec1}. GJ876 radial velocity of
$-1.3$~km/s (HARPS) implies that this shift should be $-0.9\cdot 10^{-5}$~day (this is a
``blue'' shift, since the star is approaching). The cumulative correction to $P_d$, due to
the perturbations and Roemer effect, is $-1.0\cdot 10^{-5}$~day, which is about half of
the corresponding statistical uncertainty. This correction should be added postfactum to
all estimations of $P_d$ in the fits in this paper, if such precision is necessary.

So far, we limited ourselves to the coplanar four-planet system model used by
\citet{Rivera10}. When testing more complicated models, we found that the orbital fit from
Table~\ref{tab_prelim} shows statistically significant improvement if we add a free linear
trend to the RV curve model. The estimated magnitude of this slope is about $0.18\pm
0.08$~m/(s$\cdot$yr), which results in a quite measurable RV offset of $\sim 2$~m/s,
accumulated during the observation time span. The formal significance of this trend, as
derived from the corresponding likelihood ratio statistic \citep{Baluev09a}, is about
$2.3\sigma$. Even though we have not yet took into account several important effects that
we will discuss in subsequent sections, such significance is too large to be neglected
without investigation. Initially, we interpreted this long-term slope as the geometrical
secular RV acceleration effect, mentioned by \citet{Correia10}, which should be equal to
$0.15$~m/(s$\cdot$yr) for GJ876. In fact, subtracting this predicted slope from the RV
data allows to get rid of any significant trend in the fitted model. Actually, we started
our research based on such corrected data, no longer bothering about any RV trends.
However, A.~Correia and E.~Rivera later confirmed (in private communication) that both
published RV time series already have this trend subtracted off. Therefore, we need to
find another interpretation. We can see three equally plausible sources: another
long-period unseen companion of the star, a tiny long-term instrumental drift, or some
extra errors in the RV reduction pipeline. Regardless of the actual nature of this
probable slope, we should try to take it into account, since it could significantly affect
our results. Since its source and magnitude remains a priori unclear, we add a free linear
term to the RV model. Most of our results below will refer to this model with trend,
although sometimes we will make a comparison with the trend-free model. We will also
return to a more rigorous estimation of the actual significance of this RV trend in
further sections.

\section{Correlated radial velocity jitter}
\label{sec_redj}
Let us first investigate whether the currently available RV data show some residual
periodicities, in addition to the current four-body model of the planetary system and
possible linear trend. To do this, we utilize the common periodogram-based approach with
modifications from \citep{Baluev09a}. The main modification is necessary to take into
account the likelihood function~(\ref{lik}), whereas the traditional periodograms (e.g.
the Lomb-Scargle one) implicitly utilize the $\chi^2$ function \citep{Baluev08a}. Each
value of the periodogram that we are about to use, is associated with the likelihood ratio
statistic, measuring how much our RV model improves, when we add a probe sinusoidal signal
to it. Another important modification, which we must highlight here, is that the
periodogram from \citep{Baluev09a} is not the traditionally used periodogram of the fixed
RV residuals to the base best fitting model. Rather, it requires a full re-fitting of the
whole set of the parameters and full re-evaluation of the RV residuals for each
periodogram value. This increases the sensitivity of the periodogram to faint
periodicities, since we can better model the cumulative RV variation, when the probe
signal indeed exists. We will refer to such periodogram as ``residual periodogram'', on
contrary to the ``periodogram of the residuals''. The residual periodograms are obviously
much more computationally-demanding than the traditional ones, especially when dealing
with Newtonian $N$-body fits.

\begin{figure}
 \includegraphics[width=\textwidth]{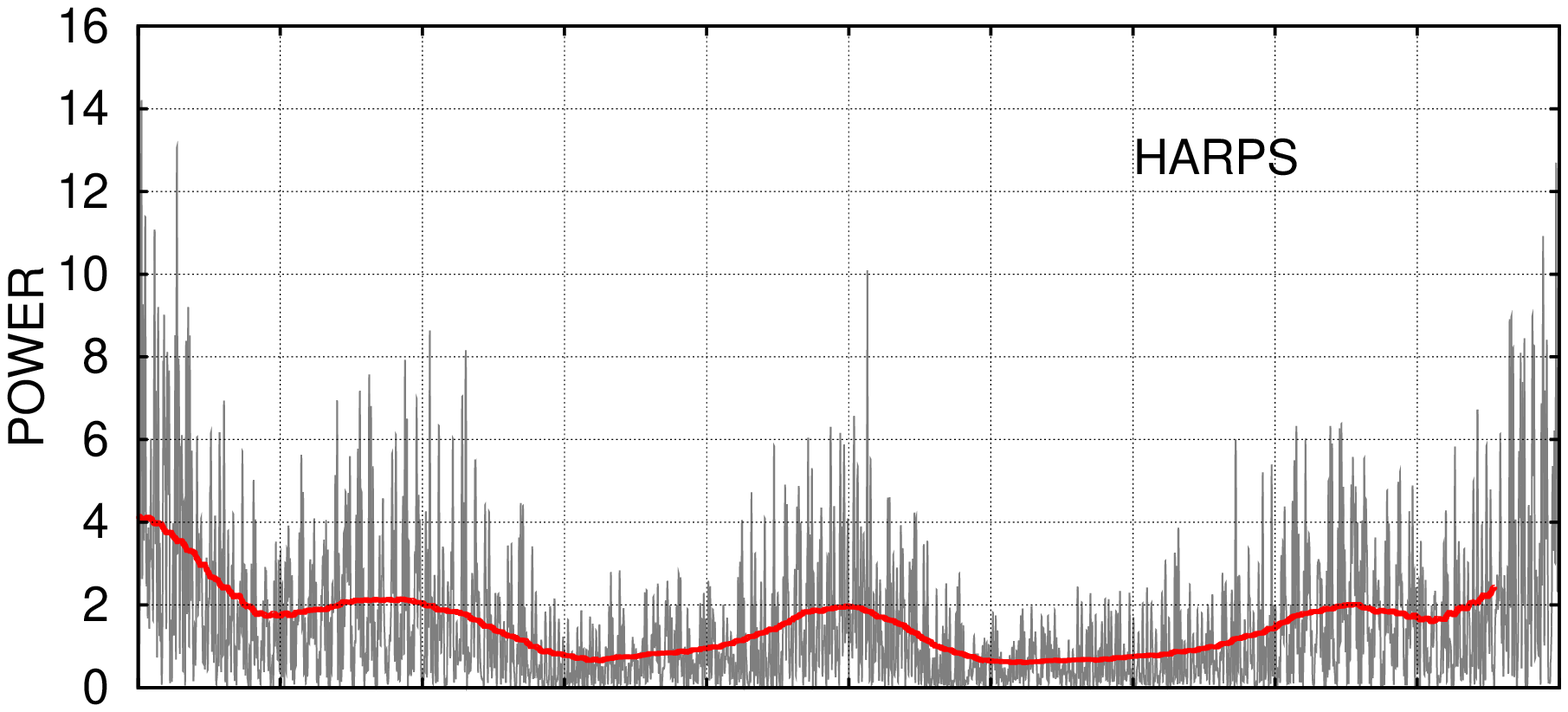}\\
 \includegraphics[width=\textwidth]{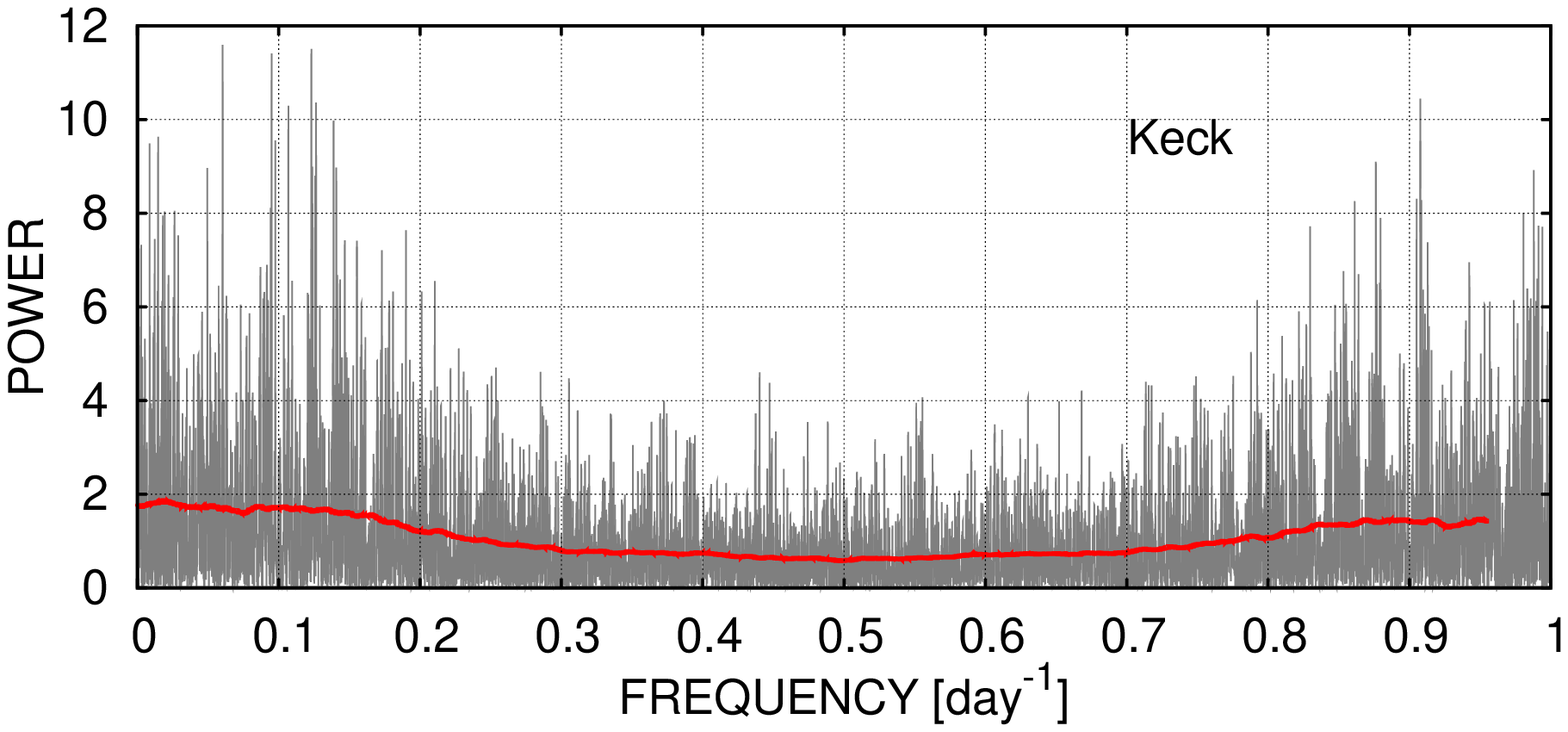}
\caption{Raw and smoothed residual periodograms for the four-planet model of GJ876 (with
eccentricity $e_e$ always fixed at zero) plus free linear RV trend. Top and bottom plots
show periodograms for the HARPS and Keck datasets. The smoothing was performed using
moving average over the frequency segments of $0.09$~\di. The periodograms (in the Keck
case especially) show excessive power at the frequencies $f\lesssim 0.2$~\di and $f\gtrsim
0.8$~\di and a relative depression in the middle of the segment.}
\label{fig_pow_linfreq}
\end{figure}

The joint residual periodogram of all available RV data, corresponding to the base model
from Table~\ref{tab_prelim} shows no isolated peaks above the apparent noise level.
However, it does not look like an usual white noise periodogram as well. On contrary, the
noise level itself demonstrates clearly varying structure. This variability becomes
especially clear, when we plot the periodogram in the linear frequency scale, instead of
the logarithmic one. Fig.~\ref{fig_pow_linfreq} shows such periodograms, plotted
separately for the Keck and HARPS datasets (i.e., assuming the probe signal is present in
only Keck or only HARPS data). The joint periodogram represents some mixture of these. We
can see that in case of the Keck data there is an excessive power at low frequencies (long
periods) and near the unit frequency (period close to one day), with a depression in the
middle of the segment. A similar frequency distribution, although somewhat obscured by
irregular variations, can be seen in the HARPS data too. These frequency spectra are very
resistant with respect to various modifications in the RV curve model. The plots from
Fig.~\ref{fig_pow_linfreq} correspond to the circular orbit of the fourth planet, but
assuming other reasonable values of $e_e$ and $\omega_e$ (see the next section) does not
significantly affect the smoothed periodograms (although individual periodogram peaks can
be affected). The influence of the long-term RV trend on these spectra looks also
insignificant, as well as the influence of possible system non-coplanarity. We could not
find a way to explain the non-uniform shape of the smoothed periodograms via any possible
shortcomings in the RV curve model. Therefore, we need to consider another explanations.

The errors in astronomical time series are usually assumed mutually uncorrelated. Such
uncorrelated sequence of errors is also known as white noise, which is called so because
of its uniform frequency spectrum. A non-uniform spectrum indicates autocorrelated
residuals, according to the Wiener-Khinchin theorem. This means, in particular, that the
fitting methods that we used above, actually are not applicable here, since they all are
based on the assumption of uncorrelated RV errors. In such a case, the correlated RV noise
could be misinterpreted as some deterministic variation, and could cause unqualified
systematic errors of the estimations in Table~\ref{tab_prelim}.

The variations of the noise level in Fig.~\ref{fig_pow_linfreq} are rather large. The
max/min ratios for the smoothed periodograms are $3.1$ and $7.4$ for the Keck and HARPS
periodograms, respectively. Nevertheless, the apparent variations of the periodogram noise
could also emerge purely by chance, because of the finite number of observations (even
when the original noise is white). To demonstrate that the variations of the average noise
level in Fig.~\ref{fig_pow_linfreq} are statistically significant indeed, and to check in
which fraction they are significant, we carry out some Monte Carlo simulations. We adopt
the fit from Table~\ref{tab_prelim} as the basic one, and carry out a series of bootstrap
simulations of the \emph{residual periodograms} plotted in Fig.~\ref{fig_pow_linfreq}. The
bootstrap random shuffling destroys any possible correlations between different residuals,
so the simulated RV noise is practically white. Therefore, the simulated periodograms
should contain only natural random variations, which we should expect for the
\emph{uncorrelated} error noise of the same variance and distribution. Each simulated
periodogram was evaluated using literally the same algorithm as real periodograms in
Fig.~\ref{fig_pow_linfreq}, and was further smoothed to access its average level. The
smoothing was done using moving average over frequency segments of $\approx 0.09$~\di. The
result of this procedure is a bunch of simulated smoothed white noise periodograms. Then
we calculate confidence ranges for the smoothed periodograms, based on the simulated
periodogram set (separate confidence range for each frequency). Also, we calculate for
each smoothed simulated periodogram the ratio of its maximum value over the whole
available frequency range to the minimum one, and count how frequently it exceeds the same
ratio of the original periodogram of the real data. It will help us to check, whether the
observed variations in the basic smoothed periodograms are statistically significant or
not.

\begin{figure}
 \includegraphics[width=\textwidth]{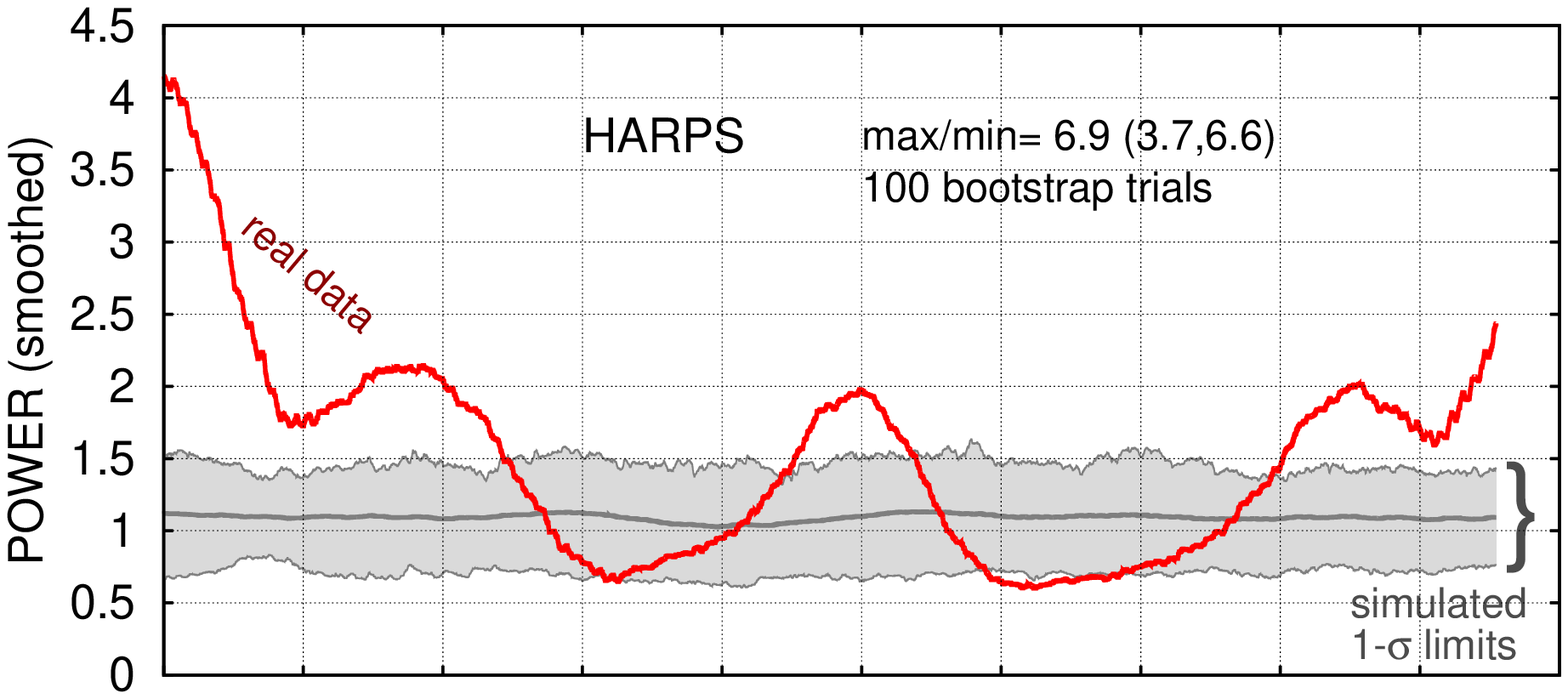}\\
 \includegraphics[width=\textwidth]{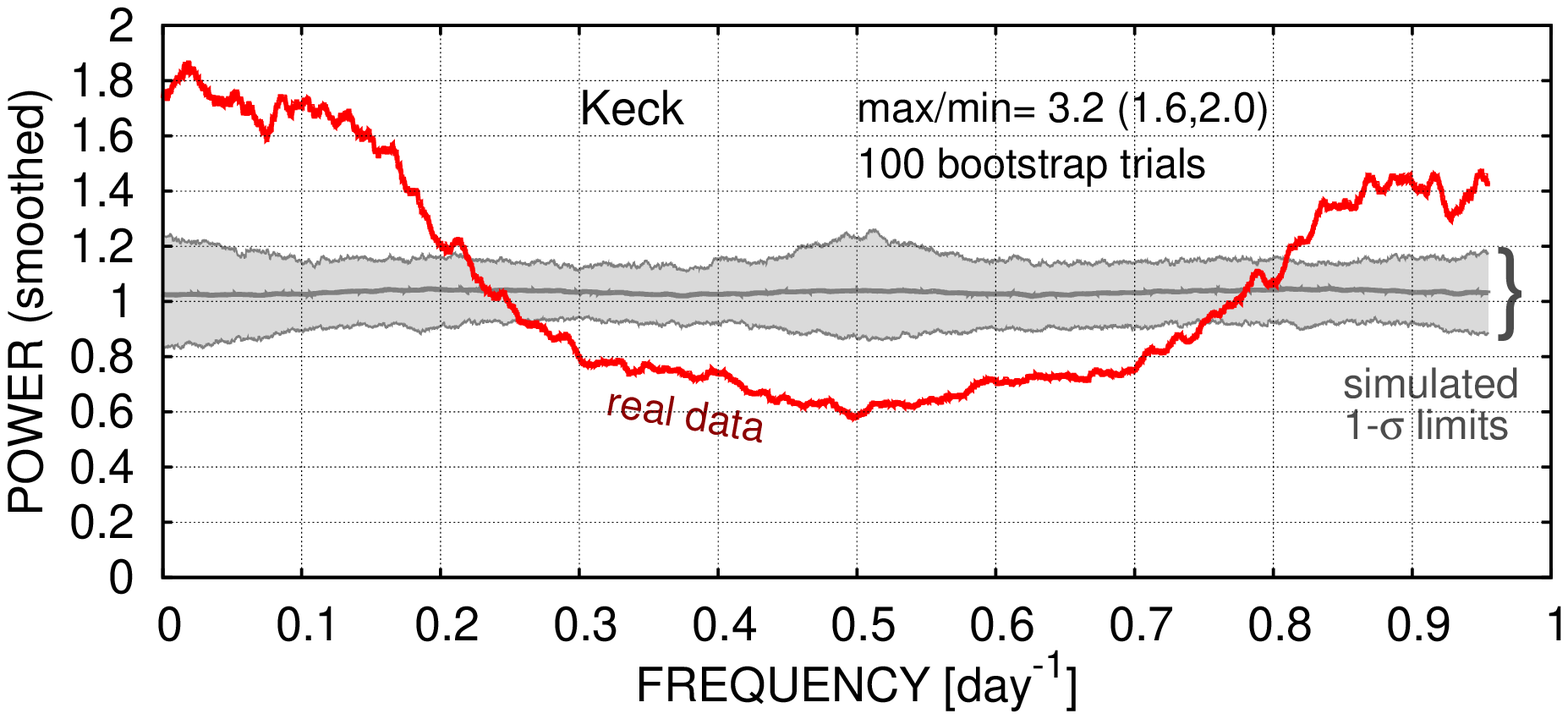}
\caption{Smoothed residual periodograms for the four-planet model of GJ876 in comparison
with their simulated levels expected for uncorrelated noise. Top and bottom plots
correspond to the HARPS and Keck datasets. The smoothing was performed using moving
average over the frequency segments of $0.09$~\di. The graphs show the actual smoothed
periodograms, their simulated mean levels and levels corresponding to the $16\%$ and
$84\%$ percentiles (two-sided one-sigma limits). The max/min ratios of the original
smoothed periodograms are also printed in each panel, along with their simulated one- and
two-sigma upper limits (in parenthesis). We can see that both Keck and HARPS data show
statistically significant deviations from the white noise in terms of their power
spectra.}
\label{fig_pow_smth}
\end{figure}

The results of these simulations are plotted in Fig.~\ref{fig_pow_smth}. We can see that
the simulated range of the smoothed periodograms remains almost constant in frequency.
This simulated range is quite narrow for the Keck case, and more wide for the HARPS case,
obviously because the number of the Keck observations considerably exceeds the number of
the HARPS ones. Both periodograms of the real data do not stay in the simulated confidence
ranges. In the Keck case, none of $100$ simulated smoothed periodograms could demonstrate
the same or larger max/min ratio as we see in the corresponding real periodogram. This
means that the noise in the Keck data is not white, with high confidence probability well
above $99\%$. In the HARPS case, the confidence probability is smaller~-- approximately
$95\%$~-- but nevertheless is high enough to say that a similar non-whiteness probably
exists in the HARPS data too, although significantly obscured by the normal random
variations.

We will not try to determine here possible physical sources for such behavior of the RV
noise, since this is an topic for another research. Regardless the actual sources, two
main practical questions arise now: how much the noise non-whiteness affects the best
fitting orbital configuration from Table~\ref{tab_prelim}, and how to correct this effect?
These problems must be solved before we can go any further. We find that the correlated
noise is already a routine issue in the exoplanetary transit searches \citep{Pont06}. The
photometric observations in these surveys often contain the so-called ``red'' noise
component. Such name is due its power spectrum, which monotonously decreases with the
frequency, thus making long-term (``red'') variations to prevail over the short-term
(``blue'') ones\footnote{There is also a \emph{narrow} notion of ``red noise'' as a
synonym of the Brownian noise, that has a specific frequency spectrum $\propto 1/f^2$. We
understand the term ``red noise'' in a general sense, however.}. Actually, the red noise
is rather common phenomenon: it is not limited to only astronomy, and it is frequently
faced in very different branches of the science.

The red noise indeed can easily explain the shape of the periodograms in
Fig.~\ref{fig_pow_linfreq}: the low-frequency ``hill'' is just directly seen in both
graphs, and another (smaller) hill near the unit frequency is its alias (caused by severe
diurnal gaps in the time series). The transit observations considered in \citep{Pont06}
allowed a simplified red noise reduction, due to their very specific distribution in time
and a specific of their photometric transit models. In fact, \citet{Pont06} could avoid,
for instance, any assumptions about the shape of the noise autocorrelation function.
Unfortunately, the approach used by \citet{Pont06} is not applicable to our case. We have
to model the correlated noise in full.

The details of the algorithm, which we use to eliminate the impact of the red RV jitter,
are given in Appendix~\ref{sec_fitcorr}. Here we only note that this algorithm is based on
a certain modification of the function~(\ref{lik}), to take the correlated noise into
account. After the maximization of this new objective function, we get (at once) the
estimations of the RV curve parameters (therefore, of the planetary orbital elements and
masses), and several RV noise parameters: different ``white'' jitters for separate time
series, $\sigma_{\mathrm{white},j}$, the common ``red'' jitter shared among different time
series, $\sigma_{\mathrm{red}}$, and the red noise correlation timescale $\tau$. Such
noise separation could be realistic if the red noise was actually caused by the star (e.g.
by some long-living photospheric phenomena) rather than by the instruments.

\section{Determinability of the fourth planet orbit}
\label{sec_fourth}
Further investigation of the likelihood function~(\ref{lik}) showed that it possesses
multiple maxima, which do not differ much from each other, in terms of the goodness-of-fit
measures like r.m.s. or the statistic $\tilde l$. Such phenomenon could indicate that the
whole orbital model is actually ill-determined due to the lack of the RV data and/or its
insufficient precision, like that was for the extrasoalar planetary system discussed in
\citep{Baluev08c}. However, it is not the case here. For the GJ876 planetary system, the
irregularities of the likelihood function are all related to only two of the free
parameters from Table~\ref{tab_prelim}, namely the eccentricity and the pericenter
argument of the fourth planet. Fixing them at some reasonable a priori defined values
makes the shape of the likelihood function very regular and practically parabolic, as it
should be when the RV model is well-linearizable with respect to the remaining free
parameters.

We check this for the following pairs of parameters: $(e_c,i)$, $(P_b,P_c)$,
$(\psi_b,\psi_e)$, and $(\omega_b,\omega_c)$. The estimations of these variables
possess the largest mutual correlations in the pairs, respectively $-0.98$, $-0.82$,
$-0.81$, and $-0.72$ (for $e_e$ fixed at zero). Correlated parameters should be usually
more affected by non-linearity effects, as discussed in \citep{Baluev08c}. For each above
pair of parameters, we plot two-dimensional contours of the function~(\ref{lik}). We
choose three contours which outline confidence regions for the selected parameter pairs,
corresponding to the asymptotic ($N\to\infty$) confidence probabilities of $1,2,3\sigma$.
The necessary confidence probabilities were calculated from the modified likelihood ratio
statistic (logarithm of the difference between the maximum and a given value
of~(\ref{lik})), as discussed in \citep{Baluev09a}. After that, we compare these
confidence regions with the results of numerical Monte Carlo simulations. For this goal,
we use the bootstrap method as described by \citet{Marcy05}. In this method the simulated
``errors'' are generated by means of the random shuffling of the RV residuals to the best
fitting RV curve model. This method of Monte Carlo simulation is rather widely used,
because it does not require any assumptions about the shape of the error distribution, and
is expected to work even for non-Gaussian errors. We introduce only two relatively
cosmetic modifications to this method. First, during the shuffling we do not mess the
residuals belonging to different datasets (we shuffle different datasets separately from
each other). Second, after each shuffling trial, we rescale the residuals according to the
total variances for the corresponding observations (which include RV jitter). We consider
only the uncorrelated noise model, since random shuffling destroys any correlations
anyway.

\begin{figure}
\begin{center}
 \includegraphics[width=0.49\textwidth]{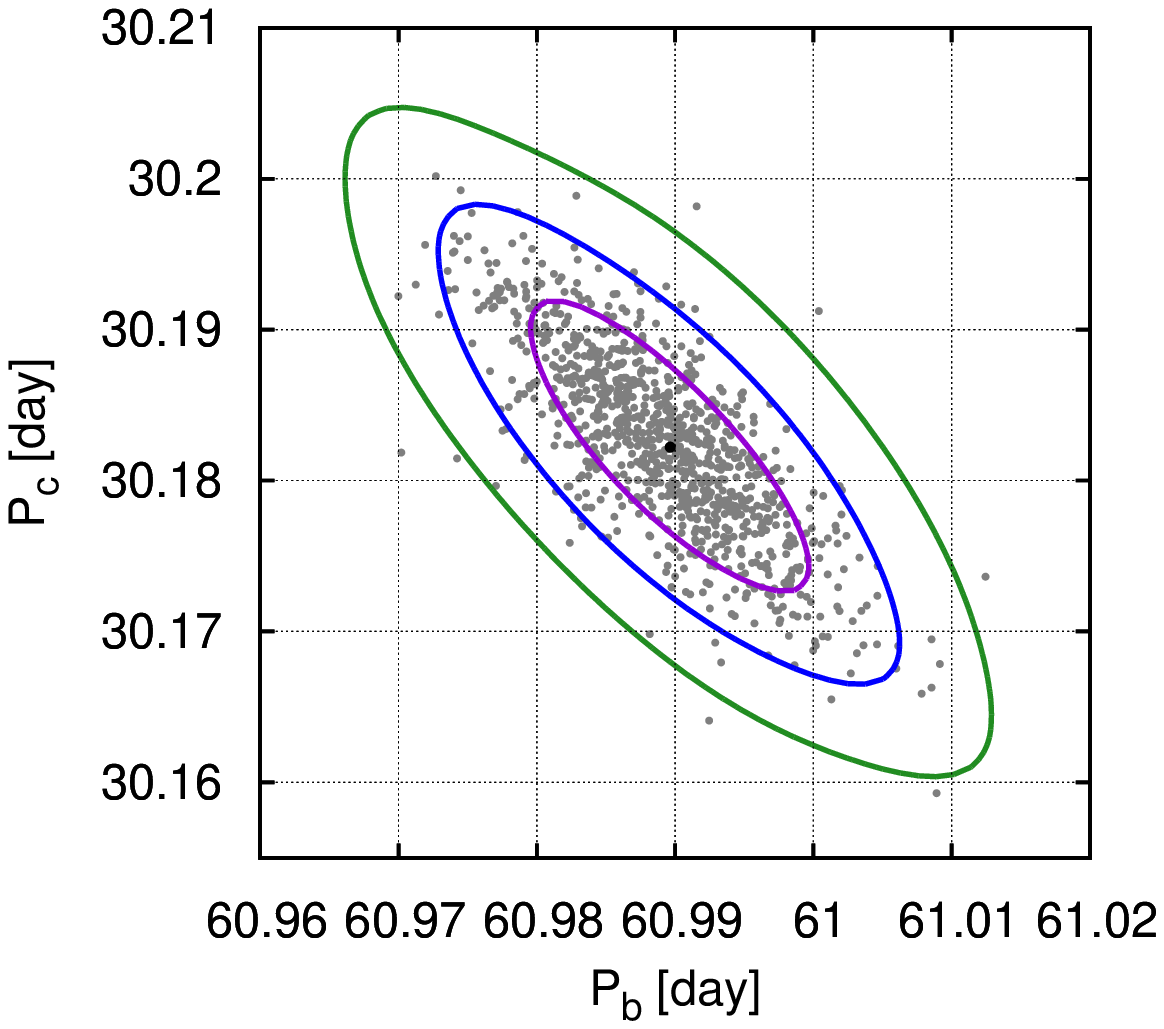}
 \includegraphics[width=0.49\textwidth]{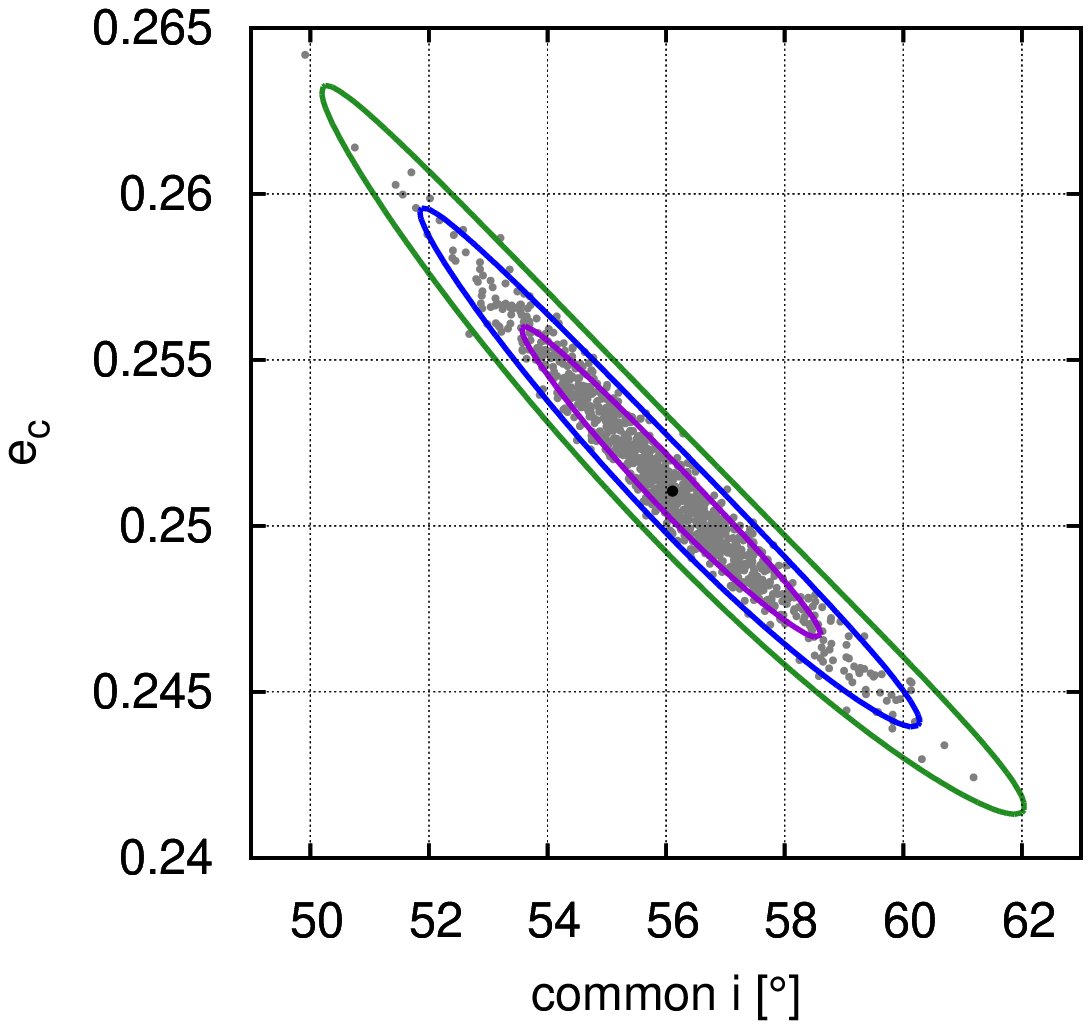}\\
 \includegraphics[width=0.49\textwidth]{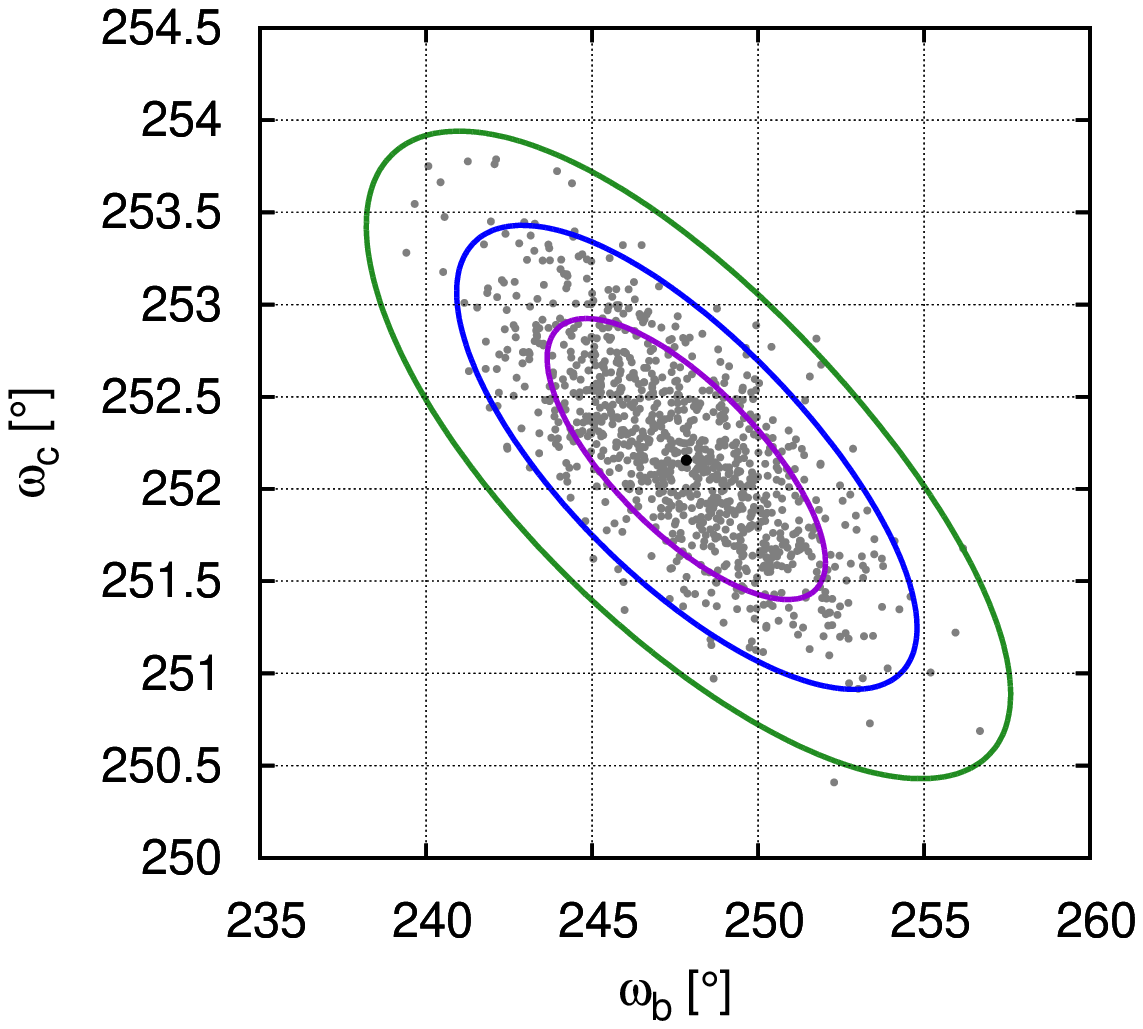}
 \includegraphics[width=0.49\textwidth]{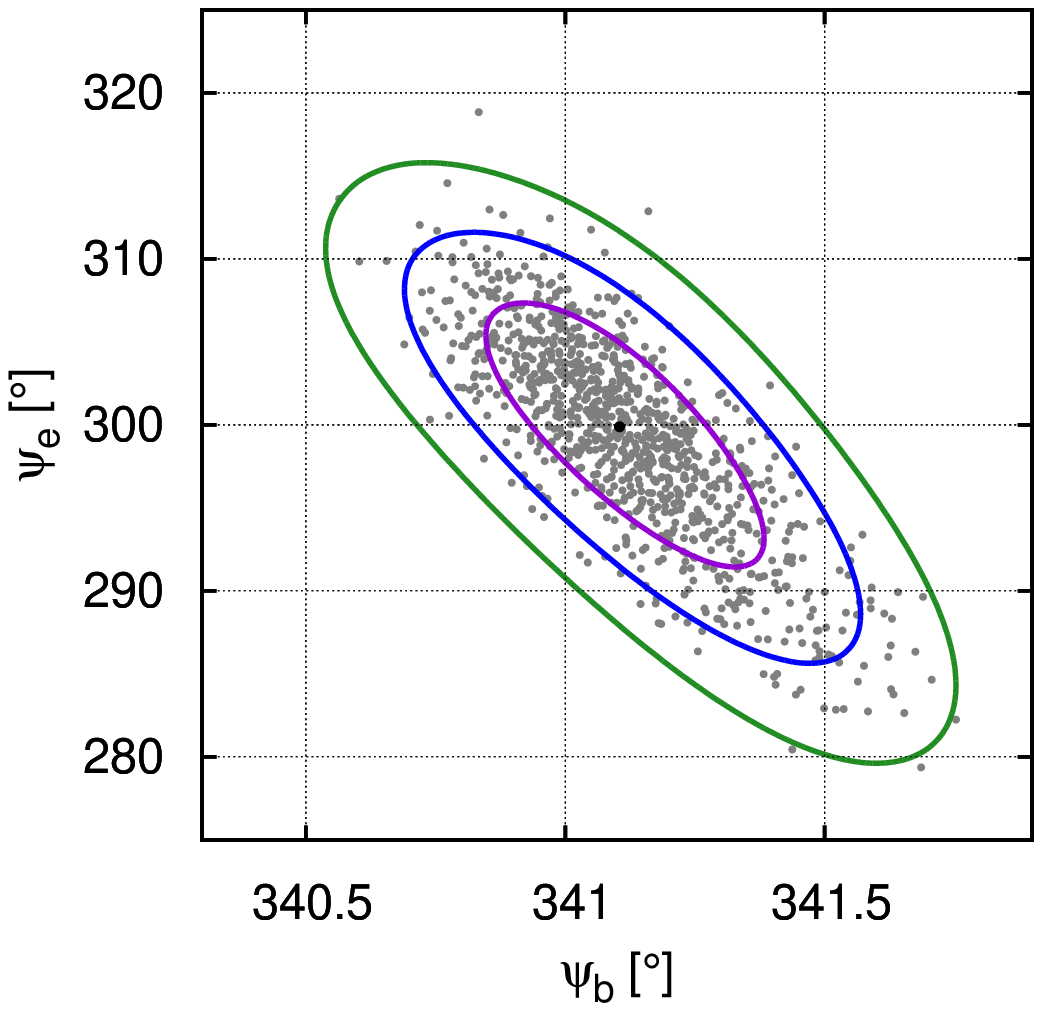}
\end{center}
\caption{Predicted confidence regions for the several selected parameter pairs of the
GJ876 planets, in comparison with numerical simulations. Each panel shows three contours
of the function~(\ref{lik}), which were calculated to outline the confidence regions
corresponding to $1,2,3$-$\sigma$ significance levels, in the framework of the linear
least-square problem (i.e., asymptotically for $N\to\infty$). These regions are all almost
elliptic and are in good agreement with results of the bootstrap simulations, shown as
points. This indicates that the problem is well linearizable with respect to the selected
parameters. The eccentricity $e_e$ was always fixed at zero here, to eliminate the
non-linearity effects coming from the $(e_e,\omega_e)$ pair. Note that these plots only
demonstrate the linearity of the selected parameters, but they should not be considered as
a source of information about the corresponding uncertainties, since many effects are not
taken into account here yet.}
\label{fig_conf}
\end{figure}

During all of the activities described in the previous paragraph, we always held $e_e$ and
$\omega_e$ fixed at zero, to eliminate all non-linearity effects associated with these
parameters. The results of these calculations are shown in Fig.~\ref{fig_conf}. We can see
that all of the resulting confidence domains have almost elliptic shape, and the simulated
sets of points are always in very good agreement with these confidence regions. This means
that when the parameters $e_e$ and $\omega_e$ are fixed, the remaining parameters behave
almost as in the linear least-squares problem: there is a single maximum of the likelihood
function~(\ref{lik}), and this function has almost parabolic shape in the vicinity around
the maximum. The estimations of the parameters possess almost Gaussian distribution, and
their uncertainty estimations are pretty reliable. Notably, even the pair involving
$\psi_e$ shows no significant non-linearity effects, despite rather large uncertainty in
this parameter.

\begin{figure}
 \includegraphics[width=0.52\textwidth]{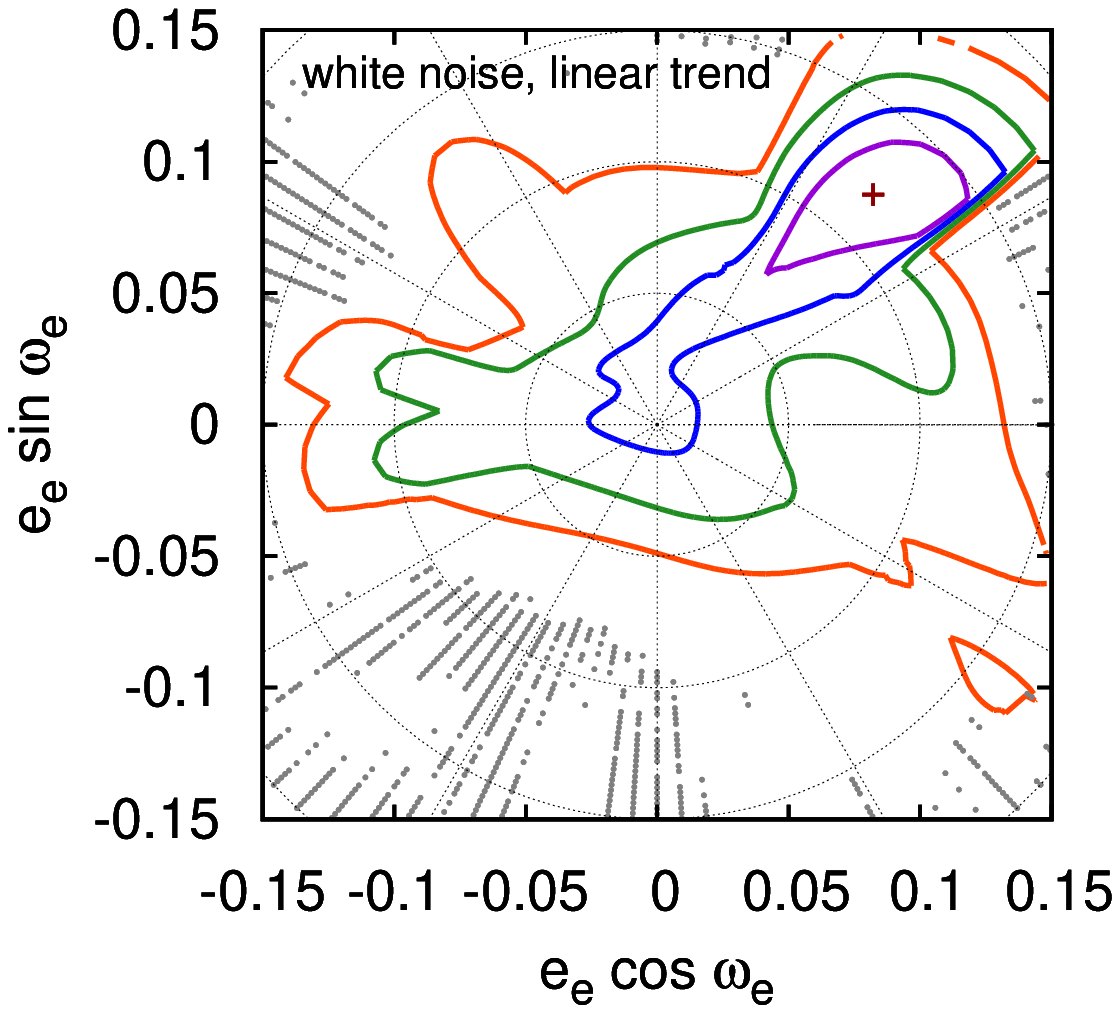}
 \includegraphics[width=0.48\textwidth]{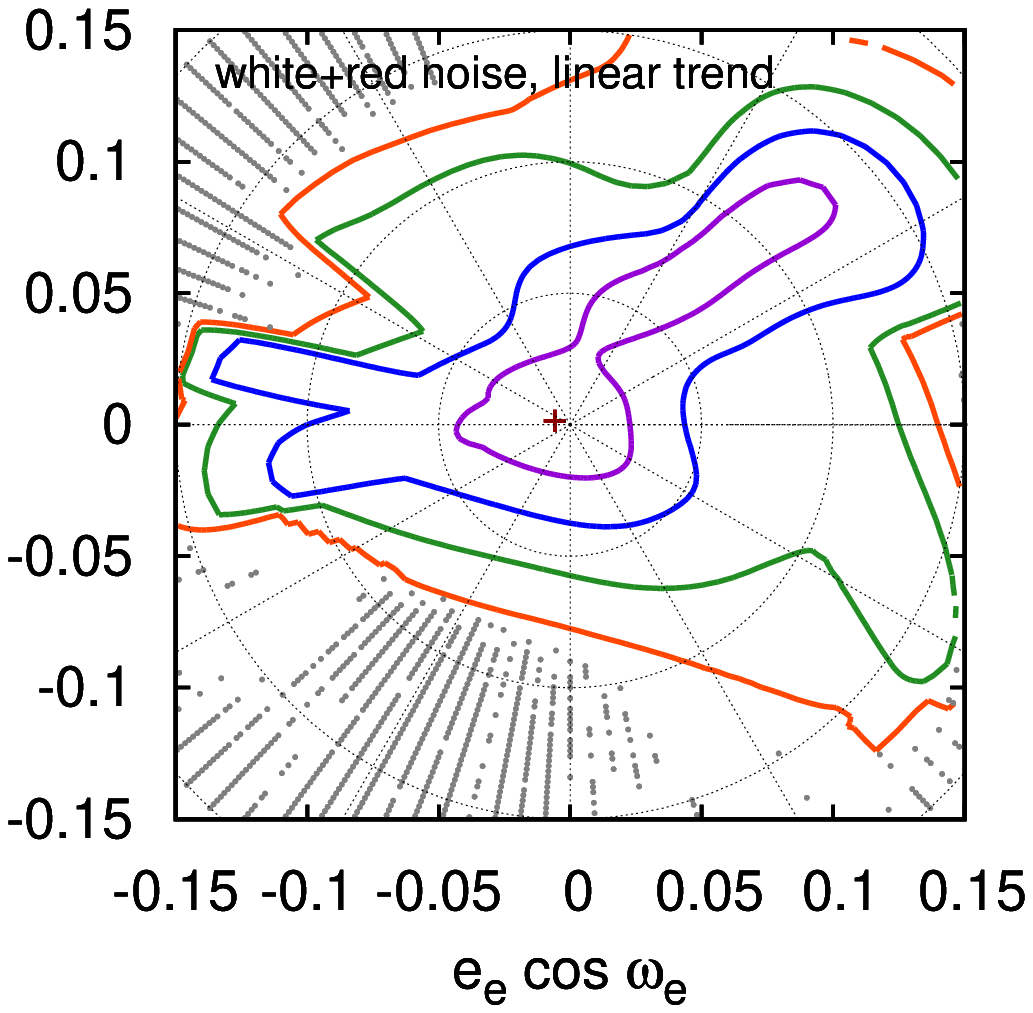}
\caption{Likelihood function contours for the parameters $(e\cos\omega,e\sin\omega)$ of
the planet GJ876~\emph{e}, assuming the RV noise is non-correlated (white) and correlated
(white+red). The gray points along radial lines mark the orbital configurations
disintegrating in less than $10000$~yrs. Each panel shows four contours of the likelihood
function, which were calculated to outline the regions of the asymptotic confidence
probability matching the $1,2,3,4$-$\sigma$ significance levels. However, the complicated
irregular and model-dependent structure of these regions indicates that they are
unreliable and may be driven by extra unqualified systematic errors in the RV time series
(see text for discussion).}
\label{fig_ew_e}
\end{figure}

The behavior of the pair $(e_e,\omega_e)$ is severely different. The similar confidence
contours for these variables look very irregular (Fig.~\ref{fig_ew_e}) and outline two
main local minima of $\tilde l$. The first local solution is the one listed in
Table~\ref{tab_prelim}, and the second one has larger eccentricity $e_e\approx 0.12$. This
second solution appears actually a bit more likely, when we use the white noise model. But
for the model with correlated noise, the first solution becomes the leading one. The
picture is also sensitive to the trend term in the RV curve model. Therefore, this
multi-extrema structure is model-dependent. Can we trust to this fine structure of the
likelihood function at all? Most probably, we cannot: it may easily change even further,
as some other non-traditional RV noise effects are discovered. The irregular details in
Fig.~\ref{fig_ew_e} are driven by RV variations at the level of $\sim 10-30$~cm/s, which
is well below the internal measurement errors ($\sim 1$~m/s). Such small variations are
hardly related to the actual radial velocity of the star. They are more likely caused by
unqualified systematic errors or systematic astrophysical noise in the data, which could
easily have amplitude about $1/3$ of the random measurement noise. Therefore, the only
reliable information that we can obtain here is that the eccentricity $e_e$ probably does
not exceed $\sim 0.15-0.20$, and $\omega_e$ values near $\sim 45^\circ$ are more favored
than the opposite ones. All other fancy details in Fig.~\ref{fig_ew_e} represent just some
misleading noise component, which we must try to eliminate.

This pseudo-detailed structure is not caused by some specific property of the fitting
method applied here. Some points in Fig.~\ref{fig_ew_e} indeed yield smaller scatter of
the RV residuals than the others, and this pattern is inevitably irregular. Any other
classic fitting method that use the function~(\ref{lik}) or any similar function as a
single goodness-of-fit measure, would yield similar results. To understand the source of
such irregular behavior of only two of the parameters, we need to trace where the
information about these bad-behaving parameters comes from. To do this, we first replotted
Fig.~\ref{fig_ew_e} for the case when the planet \emph{e} is excluded from the $N$-body RV
model, and its contribution was modeled by a Keplerian function. Such plot was already
pretty regular and elliptic, just as those in Fig.~\ref{fig_conf}. However, the
corresponding confidence domains appeared much wider than in Fig.~\ref{fig_ew_e}, allowing
$e_e$ values of up to $0.3-0.4$. This means that the contours in Fig.~\ref{fig_ew_e} could
not be settled by the non-sinusoidal shape of the RV variation contributed by the fourth
planet, as it would be in the non-perturbed Keplerian case. Instead, the information about
the likely values of $e_e$ and $\omega_e$ is extracted mainly from the perturbational
effects, which the fourth planet imposes to the motion of other planets. Therefore, the
main source of the irregularities in Fig.~\ref{fig_ew_e} is the dynamical interaction of
the planet \emph{e} with other planets in the system. This dynamical interaction allows to
considerably shrink the region of admissible values of $e_e$ and $\omega_e$, but by the
cost of considerably irregular dependence of the RV curve on these parameters. As a
result, we see some unreliable structures inside this region.

In view of the unreliable behavior of the parameters $e_e$ and $\omega_e$, the best
course of action for us will be to recognize that all points within some wide enough
contour in Fig.~\ref{fig_ew_e} are equally likely. To determine, which contour should
serve as a boundary, we apply a dynamical stability test to all orbital solutions spanning
Fig.~\ref{fig_ew_e}. Each point in this graph corresponds to some best fitting orbital
configuration of the system. For each of these configurations, we perform the numerical
integration over $10^4$~yrs and check, whether a given configuration disintegrates during
this test term. The integration term of $10^4$~yrs is actually pretty short in comparison
with cosmological timescales, but nevertheless it contains about thousand of secular
periods of the system. Therefore, it roughly corresponds to $\sim 10^7$~yrs, when rescaled
to the subsystem of the giant planets in our Solar System, for instance. Such time segment
is long enough to determine approximate stability boundaries. In Fig.~\ref{fig_ew_e}, this
stability boundary passes close to the formal $4\sigma$ confidence contour (in case of the
white+red noise model). Since the stability region should somewhat shrink, when the
integration time increases, we believe that it is safe to adopt the formal $3\sigma$
contour as a boundary outlining the admissible values of $e_e$ and $\omega_e$.

In view of such rather poor determinability of the fourth planet eccentricity, one can
ask: whether this planet exists at all? It was detected by \citet{Rivera10} on the basis
of the Keck data only, and \citet{Correia10} did not note it in their HARPS and old Keck
data. We also checked the residual periodograms for the three-planet model, and find that
the fourth planet is indeed strongly supported by the Keck data, but the HARPS periodogram
does not show a significant peak above the apparent noise level. One can suspect that the
RV signal from this fourth planet could be actually caused by some spurious drifts in the
Keck data.

\begin{figure}
\begin{center}
 \includegraphics[width=0.75\textwidth]{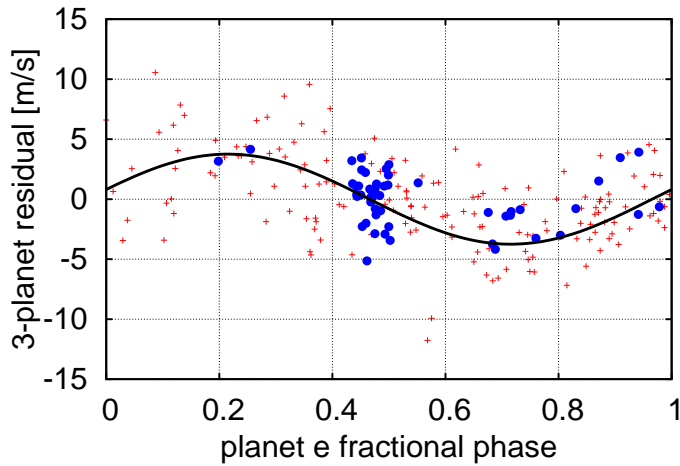}
\end{center}
\caption{Residuals of the best fitting three-planet model of GJ876, phased to the orbital
period of the fourth planet. Small crosses mark the Keck data, fat points stand for the
HARPS data. Solid curve is the unperturbed sinusoidal model of the RV oscillation due to
the planet \emph{e}.}
\label{fig_resid3}
\end{figure}

However, these doubts dissolve when we look at the RV residuals to the three-planet model
directly. We can see (Fig.~\ref{fig_resid3}), that the available HARPS data actually
\emph{confirm} the RV signal from the fourth planet and are in good agreement with the
Keck data. This agreement just is not statistically significant yet, however it may become
more significant, when more HARPS data are accumulated. At present, we have no
observational basis for doubts in the existence of the planet GJ876~\emph{e}.

\section{The reference orbital fit}
\label{sec_nom}
Using the algorithm from Appendix~\ref{sec_fitcorr}, we can now obtain the full set of the
parameters, taking into account the effect of the correlated noise. However, we must also
address the influence of the bad-behaving parameters $(e_e,\omega_e)$. Without any extra
care, we will get just a few similar orbital solutions having unrealistic uncertainties.
We adopt the following approach. Since other parameters behave almost linearly (see
Fig.~\ref{fig_conf}), we first perform a basic orbital fit, fixing the value of $e_e$ and
$\omega_e$ at some reference realistic value (say, $e_e=0$). The resulting error
estimation for each of the fitted parameters reflects only a fraction of the full
uncertainty. We also need to estimate the uncertainty caused by the non-determinability of
the parameters $e_e$ and $\omega_e$. To do this, we vary $e_e$ and $\omega_e$ inside their
admissible region (which was defined in Sect.~\ref{sec_fourth}) and see, how much this
affects the best fitting values of other (well-behaving) parameters. This gives us the
remaining part of the total parameter uncertainty (generally asymmetric).

\begin{table}
\caption{Reference orbital solution for the GJ876 system, assuming only white RV noise
(epoch JD2452000).}
\begin{center}
\begin{tabular*}{\textwidth}{@{}ll@{\kern1mm}l@{\kern1mm}l@{\kern1mm}l@{}}
\hline\noalign{\smallskip}
parameter            & planet b (*)                & planet c (*)                  & planet d                  & planet e (*)                 \\
\multicolumn{5}{c}{fitted planetary parameters} \\
$P$~[days]           & $60.990\err{6}{20}{30}$     & $30.182\err{6}{9}{16}$        & $1.937888\err{18}{6}{6}$  & $124.5\err{0.4}{5.4}{1.6}$   \\
$\tilde K$~[m/s]     & $213.28\err{33}{21}{29}$    & $84.48\err{35}{76}{61}$       & $6.20\err{28}{14}{21}$    & $3.6\err{0.3}{0.2}{1.0}$     \\
$\psi$~[$^\circ$]    & $341.11\err{19}{51}{62}$    & $72.0\err{0.4}{1.1}{0.7}$     & $357.5\err{3.1}{0.5}{0.5}$& $300\err{6}{43}{28}$         \\
$e$                  & $0.0332\err{13}{17}{38}$    & $0.2511\err{28}{63}{42}$      & $0.148\err{45}{16}{25}$   & $0(<0.19)$                   \\
$\omega$~[$^\circ$]  & $247.8\err{2.7}{9.3}{5.3}$  & $252.2\err{0.5}{0.6}{1.6}$    & $217\err{19}{14}{14}$     & $0$(unconstr.)               \\
$i$~[$^\circ$]       & \multicolumn{4}{c}{
                                          $56.1\err{1.5}{1.6}{3.8}$
                                         } \\
\multicolumn{5}{c}{derived planetary parameters} \\
$m$~[$M_{Jup}$]      & $2.39\err{4}{12}{4}$        & $0.750\err{13}{32}{12}$       & $0.0221\err{11}{13}{10}$   & $0.051\err{5}{4}{14}$        \\
$a$~[AU]             & $0.211021\err{15}{56}{40}$  & $0.131726\err{17}{28}{49}$    & $0.02110627\err{13}{6}{5}$ & $0.3397\err{7}{96}{24}$      \\
\end{tabular*}
\begin{tabular*}{\textwidth}{@{}lll@{}}
\noalign{\smallskip}\hline\noalign{\smallskip}
                     & Keck                        & HARPS                       \\
\multicolumn{3}{c}{RV data series and general fit parameters}                    \\
$c_0$~[m/s]          & $50.66\err{28}{39}{46}$     & $-1338.71\err{53}{60}{80}$  \\
$c_1$~[m/(s$\cdot$yr)] & \multicolumn{2}{c}{
                                            $0.173\err{74}{67}{86}$
                                           } \\
$\sigma_{\mathrm{white}}$~[m/s]
                     & $2.31\err{22}{19}{1}$      & $1.61\err{22}{16}{18}$      \\
r.m.s.~[m/s]         & $2.97$                     & $1.81$                      \\
\multicolumn{3}{c}{$\tilde l=5.780$/$2.764$~m/s, $d=27$}\\
\noalign{\smallskip}\hline
\end{tabular*}
\end{center}
\smallskip
The same notes as in Table~\ref{tab_prelim} apply here, except for a more complicated
treatment of the parameter uncertainties. Each estimation is now accompanied by
\emph{three} uncertainty values in the parenthesis: the first value denotes the $1\sigma$
uncertainty due to the usual linear statistical effects, assuming $e_e$ fixed at zero.
These values were calculated in the traditional way. The remaining pair of values in the
parenthesis (one above another) reflect the asymmetric uncertainty inferred by the
non-linear parameters $e_e$ and $\omega_e$. The last figure given in each uncertainty
value always maps to the last figure in the corresponding estimation. The estimations
themselves, as well as the values of the r.m.s. and $\tilde l$ are given for the basic
solution $e_e=0$. We omit the values related to the Lick, ELODIE, and CORALIE time series.
See text for further details.
\label{tab_ref}
\end{table}

\begin{table}
\caption{Reference orbital solution for the GJ876 system, with red RV noise taken into
account (epoch JD2452000).}
\begin{center}
\begin{tabular*}{\textwidth}{@{}ll@{\kern1mm}l@{\kern1mm}l@{\kern1mm}l@{}}
\hline\noalign{\smallskip}
parameter            & planet b (*)                & planet c (*)                  & planet d                  & planet e (*)                 \\
\multicolumn{5}{c}{fitted planetary parameters} \\
$P$~[days]           & $60.988\err{8}{21}{40}$     & $30.182\err{8}{12}{18}$       & $1.937879\err{17}{4}{7}$  & $124.6\err{0.5}{6.2}{1.4}$   \\
$\tilde K$~[m/s]     & $213.32\err{38}{26}{33}$    & $84.44\err{41}{81}{58}$       & $6.21\err{25}{9}{17}$     & $3.7\err{0.4}{0.0}{1.1}$     \\
$\psi$~[$^\circ$]    & $341.08\err{22}{57}{48}$    & $72.2\err{0.5}{1.2}{0.9}$     & $358.4\err{2.7}{0.4}{0.9}$& $301\err{6}{37}{27}$         \\
$e$                  & $0.0342\err{15}{38}{42}$    & $0.252\err{3}{10}{4}$         & $0.101\err{49}{22}{28}$   & $0(<0.19)$                   \\
$\omega$~[$^\circ$]  & $248\err{3}{12}{5}$         & $251.9\err{0.6}{1.0}{2.0}$    & $241\err{26}{14}{26}$     & $0$(unconstr.)               \\
$i$~[$^\circ$]       & \multicolumn{4}{c}{
                                          $55.1\err{1.8}{1.9}{4.3}$
                                         } \\
\multicolumn{5}{c}{derived planetary parameters} \\
$m$~[$M_{Jup}$]      & $2.42\err{5}{15}{5}$        & $0.759\err{15}{40}{15}$      & $0.0224\err{10}{15}{8}$    & $0.054\err{6}{3}{16}$        \\
$a$~[AU]             & $0.211025\err{19}{42}{57}$  & $0.131726\err{22}{36}{51}$   & $0.02110621\err{12}{5}{3}$ & $0.3398\err{9}{110}{26}$     \\
\end{tabular*}
\begin{tabular*}{\textwidth}{@{}lll@{}}
\noalign{\smallskip}\hline\noalign{\smallskip}
                     & Keck            & HARPS           \\
\multicolumn{3}{c}{RV data series and general fit parameters}\\
$c_0$~[m/s]          & $50.79\err{33}{30}{45}$     & $-1338.61\err{63}{45}{83}$  \\
$c_1$~[m/(s$\cdot$yr)] & \multicolumn{2}{c}{
                                            $0.223\err{98}{96}{94}$
                                           } \\
$\sigma_{\mathrm{white}}$~[m/s]
                     & $1.31\err{41}{36}{9}$      & $0.49\err{54}{34}{9}$      \\
$\sigma_{\mathrm{red}}$~[m/s]
                     & \multicolumn{2}{c}{
                                          $1.84\err{28}{37}{16}$
                                         }\\
$\tau_{\mathrm{red}}$~[day]
                     & \multicolumn{2}{c}{
                                          $3.0\err{1.7}{4.8}{0.8}$
                                         }\\
r.m.s.~[m/s]         & $2.99$          & $1.91$          \\
\multicolumn{3}{c}{$\tilde l=5.577$/$2.618$~m/s, $d=27$}\\
\noalign{\smallskip}\hline
\end{tabular*}
\end{center}
\smallskip
The same as in Table~\ref{tab_ref}, but assuming that the RV noise contains a common red
component shared between different time series.
\label{tab_refred}
\end{table}

The results of these calculations are given in Table~\ref{tab_ref} (white noise model) and
Table~\ref{tab_refred} (white+red noise model). Note that when we varied $e_e$ and
$\omega_e$, this mainly affected the estimations of other parameters themselves, and the
estimations of the statistical uncertainties remained almost constant. The first
(probabilistic) uncertainties for the most of the parameters in
Tables~\ref{tab_ref},\ref{tab_refred} should be quite reliable, with a very few exceptions
though. These exceptions are caused, however, by the statistically unsuitable non-linear
parametrization of the planetary model, rather than by the internal non-linearity of the
problem itself. The estimations of bounded parameters, like the eccentricity or the RV
jitter (both $\geq 0$) are considerably non-Gaussian and asymmetric, if the formal
uncertainty range can cover the forbidden values. It is nonetheless very easy to find a
better (more linear) parametrization in such cases. In particular, it is better to
consider the pair $(e_d\cos\omega_d, e_d\sin\omega_d)$ instead of $(e_d,\omega_d)$, as it
will be done below. When the value of some RV jitter is close to zero (in comparison with
its uncertainty) then it is better to deal with the squared jitter instead, with the
uncertainty properly rescaled (see Table~1 by \citet{Baluev09a}).

\begin{figure}
 \includegraphics[width=0.52\textwidth]{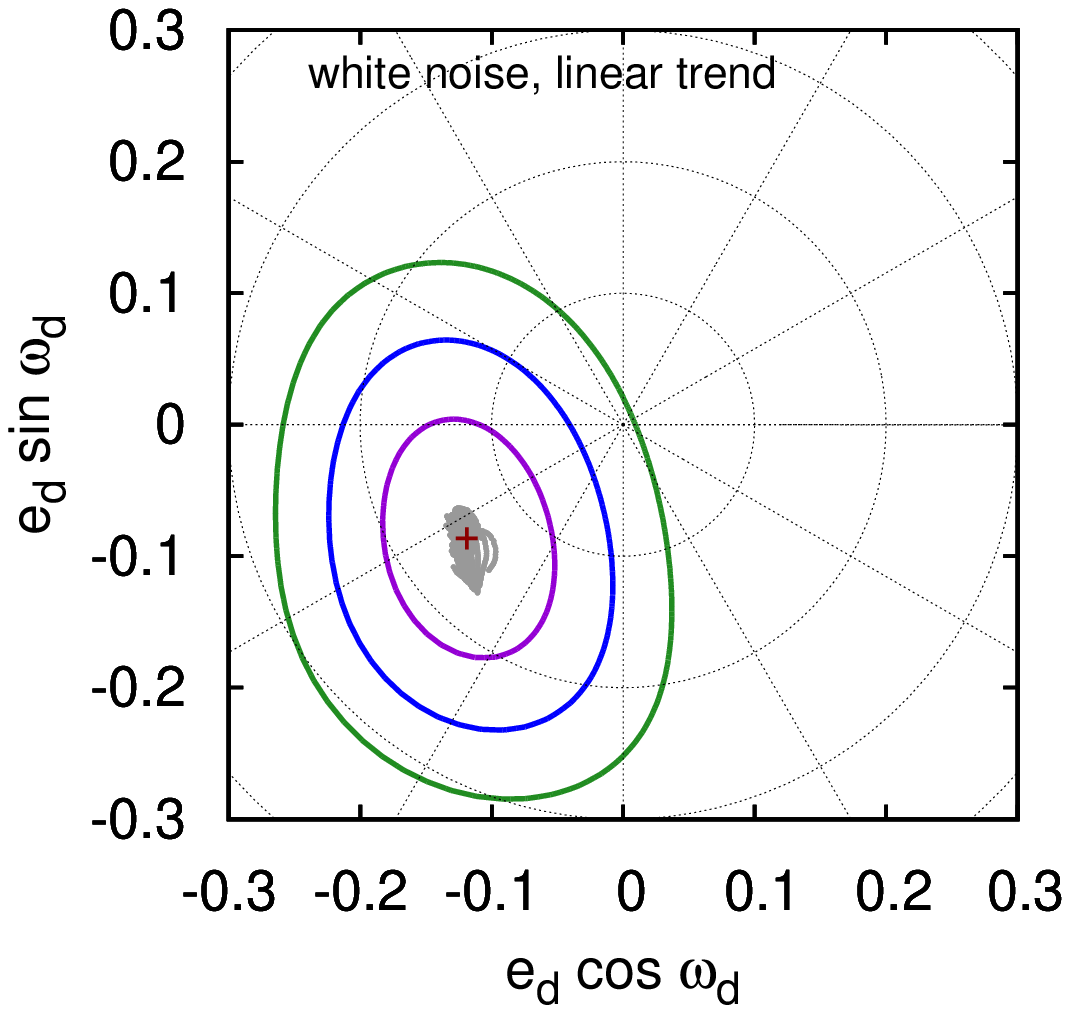}
 \includegraphics[width=0.48\textwidth]{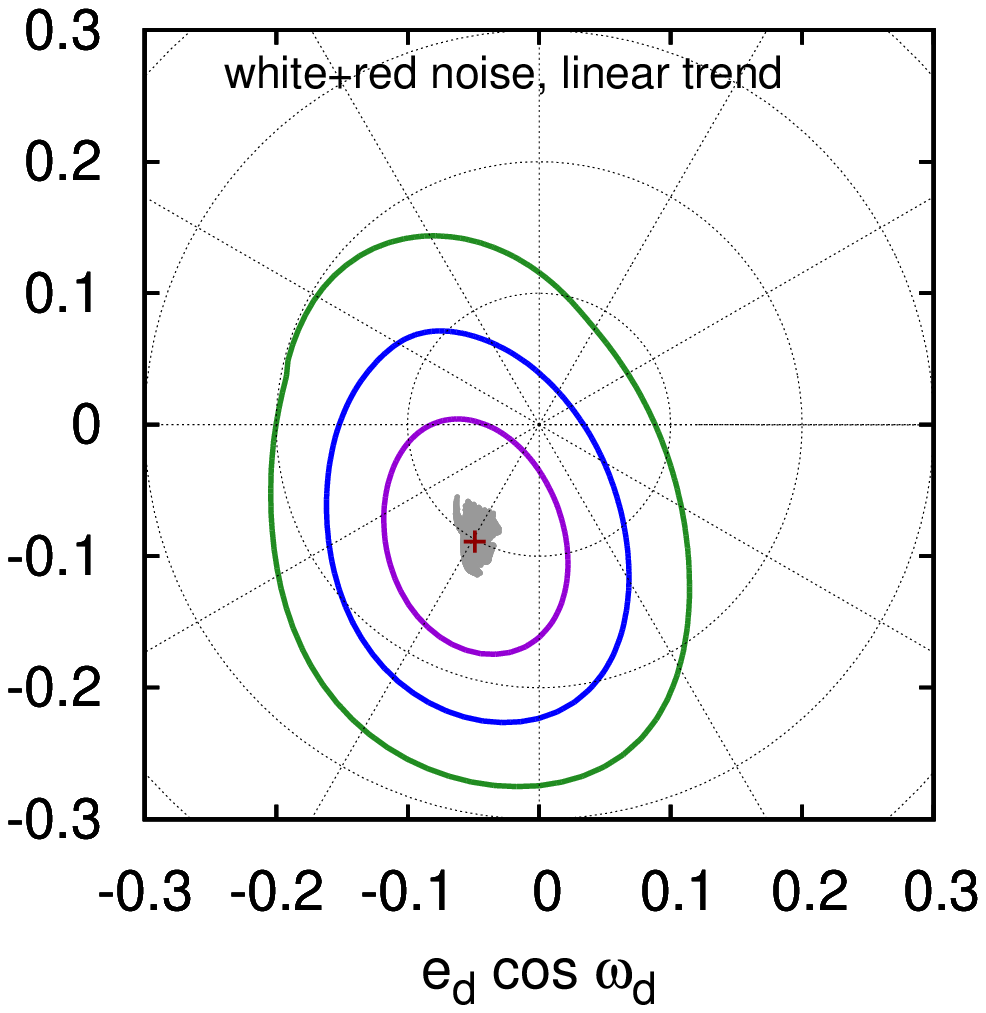}
\caption{The confidence regions for the eccentric parameters of the planet \emph{d},
plotted in the same way as in Fig.~\ref{fig_ew_e}. The stability test integrations were
not carried out here, because these parameters are not crucial for the system stability.
The confidence contours (thick lines) were plotted for the $1,2,3$-$\sigma$ probabilities
and assuming $e_e$ is fixed at zero. The remaining uncertainties, inferred by the
bad-behaving parameters $e_e$ and $\omega_e$, are rendered as gray domains around the best
fitting points for $e_e=0$ (marked as crosses). See text for discussion.}
\label{fig_ew_d}
\end{figure}

Looking at these orbital fits, we note, at first, that the linear RV trend appears more
disputable than it seemed before: some values of $e_e$ and $\omega_e$ infer too low
significance for $c_1$ (about $1-1.5\sigma$). Basically, the need for this trend can be
eliminated by means of choosing appropriate values for $e_e$ and $\omega_e$ (namely, in
the right and top regions in Fig.~\ref{fig_ew_e}). On contrary, it is too early to claim
that this trend does not exist at all. Other possible values of $e_e$ still require this
RV trend. Allowing non-coplanar configurations (see Sect.~\ref{sec_inc}) keeps this RV
trend almost intact, so it is not easy to explain this RV trend via orbital
non-coplanarity as well. We conclude that this issue can be resolved by future
observations.

Second, the red noise correlation timescale $\tau_{\mathrm red}$ is in fact poorly
constrained. We can only say that $\tau_{\mathrm red}$ has an order of days. On contrary,
the magnitude of the red jitter itself, $\sigma_{\mathrm red}$, is rather well constrained
and is well separated from zero. This suggests that although the very existence of the red
noise in the data looks supported, the characteristics of this jitter are still difficult
to assess. Nevertheless, at present even a rough estimation of $\tau_{\mathrm red}$ can
provide important physical information.

Third, we can see some drop in the eccentricity $e_d$, after we take the red noise into
account. Since the apparently eccentric orbit of the planet \emph{d} has an important
value for, e.g., the star-planet tidal interaction theory \citep[e.g.][]{FerrazMello08},
we should investigate the parameters $e_d$ and $\omega_d$ more closely.
Fig.~\ref{fig_ew_d} contains the confidence contours for the parameters
$(e_d\cos\omega_d,e_d\sin\omega_d)$, constructed for $e_e$ fixed at zero, like in
Fig.~\ref{fig_conf}. We can see that they are perfectly elliptic and, in the white noise
case, also agree with the bootstrap simulations (not shown).

Before drawing any conclusions, we need to characterize the uncertainty coming from the
parameters $e_e$ and $\omega_e$. Again, we adopt the $3\sigma$ contour in
Fig.~\ref{fig_ew_e} as a safe region of admissible values for $e_e$ and $\omega_e$, not
paying attention to the apparent concentrations inside this region. Each point in this
region refers to some orbital fit with fixed $e_e$ and $\omega_e$. The values of $e_d$ and
$\omega_d$ from these fits would mark the centers (best fit points) of the error ellipses
in the plane $(e_d\cos\omega_d, e_d\sin\omega_d)$, if we actually constructed such
confidence ellipses for each admissible $(e_e,\omega_e)$ value. For instance, two best fit
points, shown as a crosses in Fig.~\ref{fig_ew_d}, correspond to the central points
$e_e=0$ in Fig.~\ref{fig_ew_e}. Furthermore, instead of this single point in each panel,
we can construct the full set of points $(e_d\cos\omega_d, e_d\sin\omega_d)$, which are
mapped from all admissible values of $e_e,\omega_e$. These sets are rendered in
Fig.~\ref{fig_ew_d} as gray domains around the basic best fit points. These domains
reflect how much the centers of the error ellipses in Fig.~\ref{fig_ew_d} may shift, while
the values of $e_e$ and $\omega_e$ are varied inside the admissible region. Although these
centers may shift, it turns out that the shape and size of the elliptic contours in
Fig.~\ref{fig_ew_d} remain fairly constant for different $e_e$ and $\omega_e$. The picture
only shifts in the fairly solid way. This means that the uncertainties inferred by the
bad-behaving (i.e., non-linear) parameters $(e_e,\omega_e)$ are well-separable from the
uncertainties inferred by other (well-linearizable) parameters.

Therefore, we may treat now that the resulting uncertainty region in the plane of
$(e_d\cos\omega_d, e_d\sin\omega_d)$ is constituted in a cumulative manner from the usual
probabilistic confidence domains, inferred by the linear estimation theory with $e_e$
fixed at zero (shown as regular elliptic contours), and from the uncertainty domain
inferred by the non-linear parameters $e_e$ and $\omega_e$ (shown as less regular small
gray regions). We consider that all admissible values of $e_e$ and $\omega_e$ are equally
possible, and therefore the gray uncertainty regions in Fig.~\ref{fig_ew_d} should be
understood as structureless solid entities, which just indicate how the original
probabilistic confidence regions should be bloated to take into account the uncertainty
coming from $e_e,\omega_e$.

We can see from Fig.~\ref{fig_ew_d} that $e_d$ is indeed inconsistent with zero, when we
analyze RV data assuming traditionally that the RV noise is white. The significance of
this non-zero $e_d$ is well above the $2\sigma$ level, even when we take into account the
uncertainty inferred by the non-linear parameters $e_e$ and $\omega_e$. However, it is now
obvious that this apparently significant value of $e_d$ is likely a result of the
misinterpreted red RV noise. When the red noise is taken into account, the best fitting
value of $e_d$ moves closer to zero, and simultaneously the uncertainty regions expand.
Taking into account the uncertainty coming from $e_e$ and $\omega_e$, we realize that
$e_d$ is in fact consistent with zero at the significance level of hardly above $1\sigma$.
The things remain similar when our RV curve model does not contain the linear RV trend. In
that case, $e_d$ is non-zero at $>3\sigma$ level for the white noise model, but for the
correlated model this significance drops to the same $\sim 1\sigma$ level. Although we
still cannot retract the values of $e_d$ as large as $\sim 0.15$, we nevertheless have no
observational evidences that $e_d$ is actually non-zero. Further observations can
eventually solve this question for sure, but at present it is too early to claim that the
non-zero eccentricity $e_d$ was confirmed. The apparent non-zero value of $e_d$ reported
in previous works represents basically an effect of misinterpreted red noise in the RV
data.

\section{System non-coplanarity}
\label{sec_inc}
One of the primary goals of this paper was to characterize the mutual non-coplanarity
between planets \emph{b} and \emph{c} or at least to put some limit on it. In the
non-coplanar case, we have two separate variables for the osculating inclinations $i_b$
and $i_c$, and also a pair of extra parameters determining the orientation of their
ascending nodes, $\Omega_b$ and $\Omega_c$. Two latter parameters, however, cannot be
estimated independently, because the problem is invariable with respect to arbitrary
rotation around the line of sight. In fact, we can determine only the difference
$\Delta\Omega_{bc}=\Omega_c-\Omega_b$. The main quantity that we are interested in, in
view of the orbit non-coplanarity, is the mutual orbital inclination $I$ for the planets
\emph{b} and \emph{c}, which is a function of $i_b$, $i_c$, and $\Delta\Omega_{bc}$.

We must note that during this work we will deal, most probably, with small values of $I$,
comparable to its statistical uncertainty. Such situation reveals many practical
difficulties, because our estimations of $I$ will have significantly non-Gaussian
distribution. Similar troubles arise for planetary eccentricities, when they are small
enough (comparable to their uncertainties). In the both situations, the issue is rather
formal, however. The non-Gaussian behavior of such estimations is caused by the bad choice
of the parametrization, rather than is inferred by the RV model itself. In case of the
eccentricity, the problem is usually eliminated if we consider, instead of the
eccentricity, the \emph{pair} of parameters $(e\cos\omega,e\sin\omega)$. Estimations of
these parameters usually have almost Gaussian bivariate distribution, even when $e$ is
small. All apparent problems in this case are caused by the trivial singularity of the
polar coordinate system $(e,\omega)$ in the point $e=0$. It is very likely that the mutual
inclination $I$ should be affected by a similar singularity at $I=0$, which could be
eliminated by means of transition to the variables like $(\sin I\cos\Phi,\sin I\sin\Phi)$,
where $\Phi$ is an extra auxiliary variable, determining some extra orientation angle
related to the mutual orbital inclination. Note that the variables $I$ and $\Phi$ should
be independent in the sense that for any pair of $I\in [0,\pi]$ and $\Phi \in [0,2\pi]$
there should exist a valid orbital configuration. Due to that requirement, we cannot
choose $\Phi$ to be equal to, e.g., the angle between the planetary ascending nodes
$\Delta\Omega$. For any $I<\pi/2$, the value of $\Delta\Omega$ cannot exceed $I$, and such
non-constant constraint will imply nothing except extra difficulties. We could use some
orientation angle in the Laplace plane of the system. Such choice would be physically
justified and also symmetric with respect to the two planets involved, but it would mess
the geometric parameters (like inclinations) with planetary masses, which is not
desirable.

\begin{figure}
\begin{center}
 \includegraphics[width=0.85\textwidth]{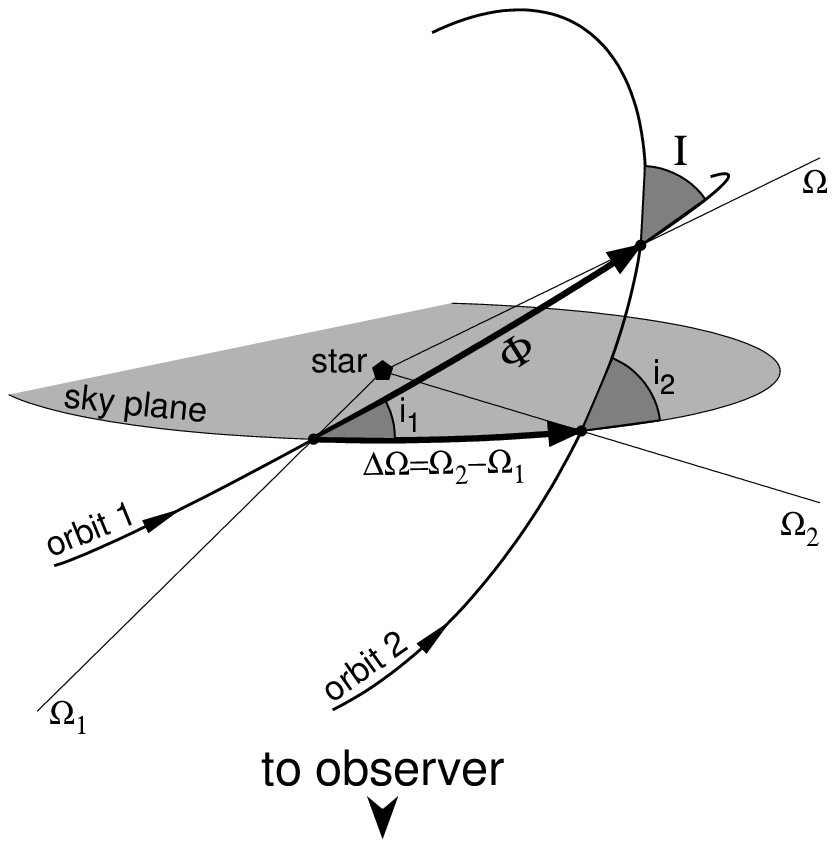}
\end{center}
\caption{Illustration of the angle $\Phi$ definition. For simplicity, both orbits are
assumed circular and of the same radius. The direction $\Omega$ is determined by the
orbits intersection point where the second planet \emph{ascends} over the first orbital
plane (and not the opposite intersection point). The directions $\Omega_1$ and $\Omega_2$
are defined in the similar way: each planet should \emph{ascend} over the sky plane in the
point where the inclination angle is determined. Such definitions allow for all
inclinations $i_1,i_2,I$ to be always non-negative (keeping implicitly their signs in the
orientation angles $\Omega_1,\Omega_2,\Phi$). See text for the detailed discussion.}
\label{fig_incl}
\end{figure}

We adopt the following purely geometric definition of the auxiliary angle $\Phi$. Given
two abstract ``first'' and ``second'' planets, we can determine two mutual orbital nodes
of these planets. We choose the node in which the \emph{second} planet \emph{ascends} over
the orbital plane of the \emph{first} planet, and define $\Phi$ as the orientation angle
of this node in the plane of the first orbit, counting it from the usual ascending node of
the first orbit. This definition is schematically illustrated in Fig.~\ref{fig_incl}. With
a help of the classical spherical trigonometry, it is not hard to derive the formulae
expressing the new osculating angles $I,\Phi$ via the original parameters $i_1,i_2,
\Delta\Omega=\Omega_2-\Omega_1$. They look like:
\begin{eqnarray}
\sin I \cos \Phi &=& - \sin i_1 \cos i_2 + \cos i_1 \sin i_2 \cos \Delta\Omega, \nonumber\\
\sin I \sin \Phi &=& \sin i_2 \sin \Delta\Omega, \nonumber\\
\cos I           &=& \cos i_1 \cos i_2 + \sin i_1 \sin i_2 \cos \Delta\Omega.
\label{IncPhi}
\end{eqnarray}
The inverse transition is given by the equalities:
\begin{eqnarray}
\sin i_2 \cos \Delta\Omega &=& \sin i_1 \cos I + \cos i_1 \sin I \cos \Phi, \nonumber\\
\sin i_2 \sin \Delta\Omega &=& \sin I \sin \Phi, \nonumber\\
\cos i_2                   &=& \cos i_1 \cos I - \sin i_1 \sin I \cos \Phi.
\label{IncPhi_bck}
\end{eqnarray}

Obviously, it is impossible to express all three original orientation parameters via only
$I$ and $\Phi$. We must also know $i_1$ to obtain $i_2$ and $\Delta\Omega$. Therefore,
such definition of $\Phi$ is not symmetric with respect to the two planets: the first
orbit serves as a reference plane. In case of GJ876, we choose the planet \emph{b} to be
this ``first'' planet, since its RV amplitude is considerably larger, and thus its orbital
plane orientation is determined with better precision. The planet \emph{c} will be the
``second'' planet, and the orbit of the planet \emph{e} is assumed to lie in the common
Laplace plane of the system (at the epoch of osculation). The planet \emph{d} was also
assumed to move in the Laplace plane, though this assumption only affects the mass
estimation of this planet.\footnote{To make these definitions more rigorous, we must also
mention that in the Jacobi coordinate system, that we adopt here, different osculating
orbits have different reference points, so they are no longer confocal, and this is not
reflected in Fig.~\ref{fig_incl}. These small displacements do not have practical
significance, however.}

Given the formulae~(\ref{IncPhi},\ref{IncPhi_bck}), we can easily carry out
\emph{constrained} RV fits with $I$ and $\Phi$ fixed at any desired values. Technically,
such constrained fitting may be done, for instance, by means of expressing $i_2$ and
$\Delta\Omega$ via $I$, $\Phi$, and $i_1$ (thus eliminating $i_2$ and $\Delta\Omega$ from
the set of free parameters). Therefore, we can perform a series of such constrained fits
on some regular grid of $I$ and $\Phi$ and then to plot the resulting likelihood contours
in the plane, e.g., $(I\sin\Phi,I\cos\Phi)$. Such plot basically visualizes the confidence
regions for these variables, similar to those regions that we have already constructed in
Fig.~\ref{fig_conf}.

\begin{figure}
 \includegraphics[width=0.53\textwidth]{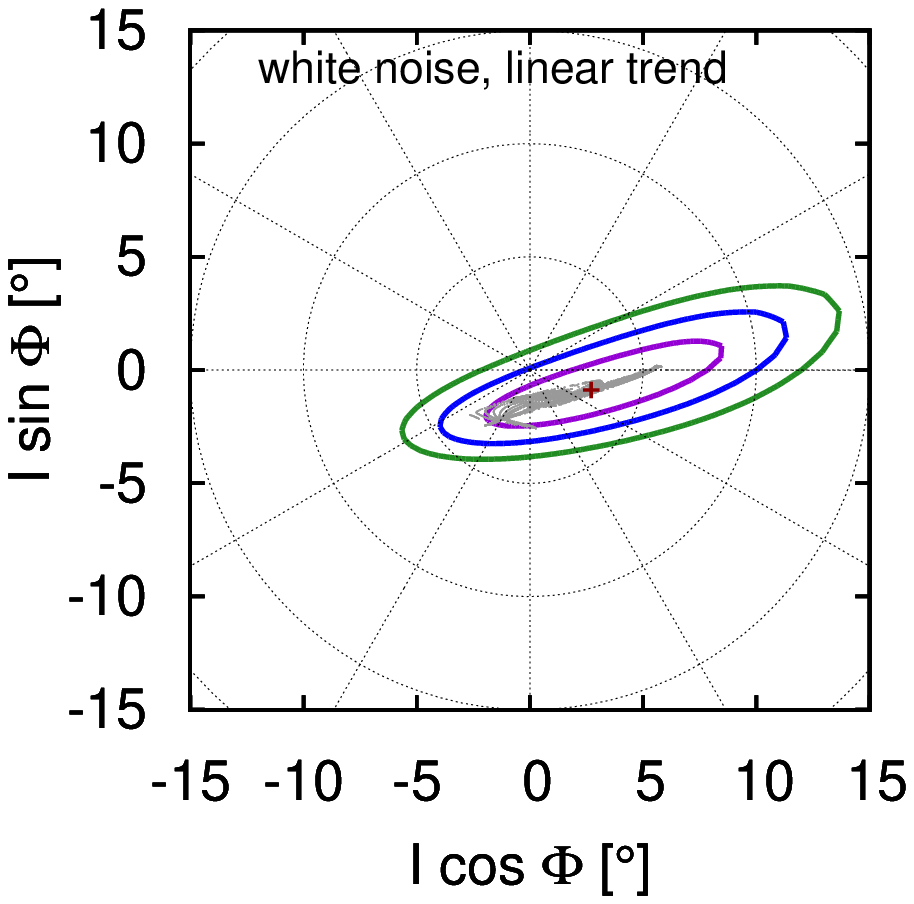}
 \includegraphics[width=0.47\textwidth]{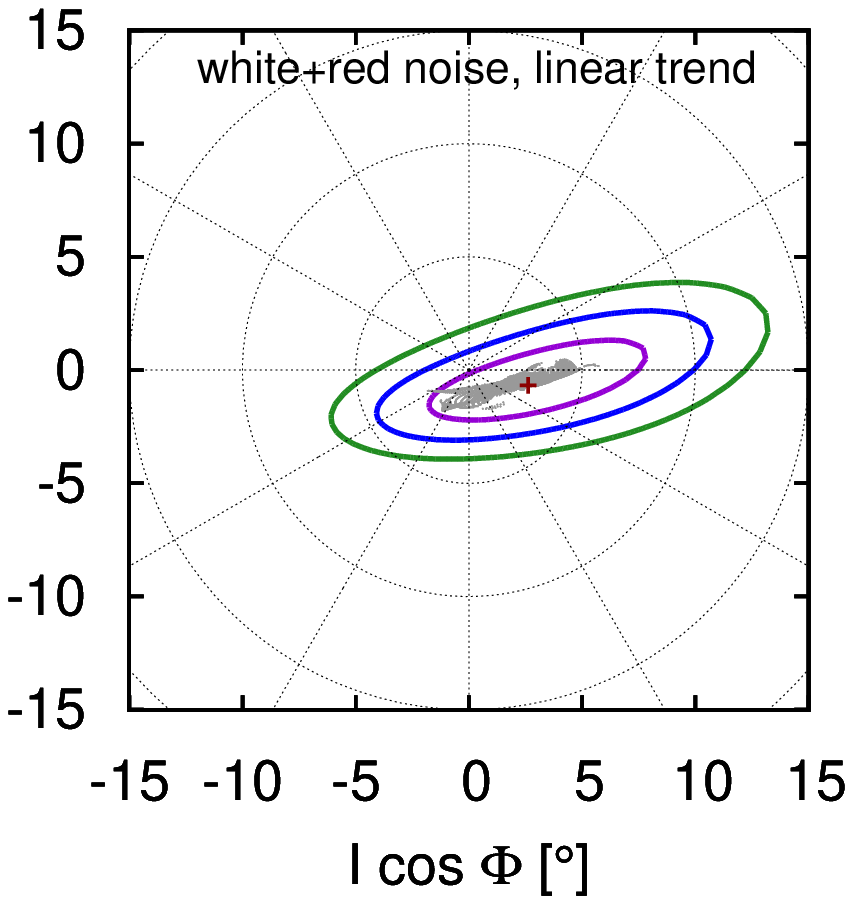}
\caption{The confidence regions for the inclinational parameters $(I\cos\Phi,I\sin\Phi)$,
plotted in the same way as in Fig.~\ref{fig_ew_d}. See text for the detailed discussion.}
\label{fig_IncPhi}
\end{figure}

Results of these calculations are shown in Fig.~\ref{fig_IncPhi} for both noise models
(white and white+red). It is notable that in the white noise case the inclination $I$
shows some non-zero value at the significance level of $1.9\sigma$, although this
significance becomes somewhat smaller when the uncertainty in $(e_e,\omega_e)$ is
included. These non-coplanarity signs are further softened, when the red noise is taken
into account. In this case, $I$ is consistent with zero at $1.2\sigma$ level, even when
the $e_e$ and $\omega_e$ uncertainties are neglected. Eventually, we conclude that
available RV data for GJ876 are fully consistent with the coplanar configuration ($I=0$).
Nevertheless, we can obtain some informative upper limit on the possible non-coplanarity
from Fig.~\ref{fig_IncPhi}: the angle $I$ likely cannot exceed $15^\circ$. However,
because of the significant prolateness of the error ellipses in Fig.~\ref{fig_IncPhi}, the
upper limit on $I$ significantly depends on the value of $\Phi$, i.e. on the orientation
of the orbital nodes for planets \emph{b} and \emph{c}. So large values of $I$ as
$10^\circ-15^\circ$ can be accepted only if the corresponding orbital planes intersect
each other rather close to the sky tangent plane (i.e., $\Phi$ close to $0$ or
$180^\circ$). When the intersection nodes are far from the sky plane, the mutual
inclination $I$ is unlikely to exceed $\sim 5^\circ$.

From the non-coplanar three-planet fit by \citet{Correia10} we find $I=1.9^\circ$ and
$\Phi=243^\circ$ from their Table~2 and $I=3.5^\circ$, $\Phi=200^\circ$ from their
Table~3. \citet{Rivera10} only mentioned that their non-coplanar four-planet fit yields
$I=3.7^\circ$. Both groups agree that there is no significant mutual inclination between
the planets \emph{b} and \emph{c}. Our results do not contradict to such estimations. We
have to admit, however, that both works discussed the non-coplanarity issue very briefly,
and they omit exact uncertainties for $I$. \citet{Correia10} gave some uncertainties for
$i_1,i_2,\Delta\Omega$ (about $1-2^\circ$), but did not supply the necessary correlations,
disabling us to derive the inferred uncertainties for $I$ and/or $\Phi$.

Previously, \citet{Bean09} also tried to determine the orbit non-coplanarity between the
planets \emph{b} and \emph{c}, based on the old Keck RV data from \citep{Rivera05} and HST
astrometry data from \citep{Benedict02}. Their non-coplanar orbital fit corresponds to
$I=4.5^\circ$ and $\Phi=330^\circ$ in our notation. These values also agree with
confidence regions in Fig.~\ref{fig_IncPhi}. We must note that although \citet{Bean09}
utilized the astrometic measurements, they, however, took into account neither the annual
systematic errors in the old Keck data (Fig.~\ref{fig_Keck_diff}), nor the very existence
of the fourth planet, nor the red noise in RV data. They also do not mention whether they
removed the secular acceleration effect from the RV data they used. Therefore, their
results cannot be directly compared with ours. Due to this, the possible effects from the
astrometry data by \citet{Benedict02} remain not fully clear. We do not expect, however,
that these effects are large. Indeed, note that the estimations of, for instance, the
absolute node longitudes $\Omega_{b,c}$ from \citep{Bean09} possess rather large
uncertainties of $\sim 8^\circ$. Since the astrometry data are responsible for these large
uncertainties almost exclusively, we may suppose that such data should not constrain very
much the values of $I\sim 5^\circ-10^\circ$, and probably most of the non-coplanarity
information now comes from radial velocities anyway. The role of the astrometric data was
nevertheless more important in \citep{Bean09}, since they used old RV data, which allowed
much larger $I$ of $\sim 15^\circ-20^\circ$. We do not use the astrometric data here,
because their error properties are in fact poorly assessed. \citet{Bean09} noted that the
actual scattering of the astrometric residuals is considerably different from the stated
instrumental uncertainties. Also, these data can easily contain a correlated component. We
think these effects are difficult to estimate reliably, due to the relatively small size
of the astrometric dataset.

In the first, purely white, case reflected in Fig.~\ref{fig_IncPhi}, the results of the
bootstrap simulations (not shown) are in good agreement with the confidence regions
plotted, just like in Fig.~\ref{fig_conf}. In the second, white+red, case, the bootstrap
simulations are not helpful, since they destroy any noise correlation effects anyway.
However, it is very likely that the formal statistical reliablity of the confidence
regions in this case should be so high as for the white noise model. The only remaining
question is how well the particular noise model, used in the algorithm from the
Appendix~\ref{sec_fitcorr}, allows to eliminate the effects coming from the RV noise
correlateness. To check this, we replotted the confidence contours from
Fig.~\ref{fig_IncPhi} assuming a bit with different noise models. First, we checked the
picture remains practically the same for different simple noise correlation functions
($e^{-|x|}$, $e^{-x^2/2}$, and $1/(1+x^2)$). After that, we probed different splitting of
the red part of the RV noise between the Keck and HARPS data. We checked the cases when
the HARPS noise is purely white, and when it has its own red component, not tied to the
Keck one. In these cases, the changes in the confidence regions are larger, roughly
similar to the difference between left and right panels in Fig.~\ref{fig_IncPhi}. This may
indicate that the effect of the red RV noise still may be not taken into account in full.
It seems from the current data, that the HARPS red jitter may be a bit smaller than the
Keck one. We cannot use a more accurate noise model, however. The Keck red RV jitter alone
looks pretty estimatable without the HARPS data, but the HARPS red jitter is, on contrary,
poorly separable. We have to live with that problem until more HARPS data are acquired.
Anyway, all noise models tested so far, do not imply a significant non-coplanarity of the
system and place almost the same limits on it.

\section{Long-term dynamics}
\label{sec_dyn}
Now we can investigate the long-term dynamical evolution of the planetary system. We
choose two best fitting orbital configurations, corresponding to the osculating $e_e=0$
(solution I) and $e_e=0.12,\omega_e=45^\circ$ (solution II). Both configurations are
obtained assuming the white+red RV noise model. We tracked the evolution of each orbital
configuration over $1$~Myr term (the innermost planet \emph{d} was not taken into account
during the integration). The relative energy error was about $10^{-8}$ in the first
integration, and about $10^{-7}$ in the second one. The energy error was walking around
these values all the time and did not show a notable accumulation effect. This is exactly
what we should expect from the symplectic integrator. We must note, however, that our
dynamical analysis here is still rather preliminary and in future it is better to  perform
the integration over longer terms and to take into account the short-period planet
\emph{d} using, e.g., an averaged Hamiltonian method \citep{Farago09}.

Both orbital configurations appeared stable during the integration time. The evolution of
the parameters related to the massive planets \emph{b} and \emph{c} is fairly regular and
is close to the apsidal corotation resonance state. The main secular period of the
system~-- the period of the pericenters revolution~-- is close to $14.3$~yrs in the both
cases. The perturbations from the fourth planet are rather small, although they add some
minor chaotic component in the motion of the massive planets. The long-term evolution of
this fourth planet itself looks, on contrary, considerably chaotic, in terms of its
orbital eccentricity at least. This eccentricity, however, remained bounded from the upper
side by $0.09$ (solution I) and $0.16$ (solution II).

We also considered the evolution of the following resonant variables, corresponding to the
individual two-planet resonances:
\begin{eqnarray}
s_{cb,c}=2\psi_b-\psi_c-\omega_c, \quad s_{cb,b}=2\psi_b-\psi_c-\omega_b, \nonumber\\
s_{be,b}=2\psi_e-\psi_b-\omega_b, \quad s_{be,e}=2\psi_e-\psi_b-\omega_e, \nonumber\\
s_{ce,c}=(4\psi_e-\psi_c)/3-\omega_c, \quad s_{ce,e}=(4\psi_e-\psi_c)/3-\omega_e.
\label{resang}
\end{eqnarray}
These definitions of resonant angles are derived from the general definitions from
\citep{Beauge03}. Since we consider only coplanar configurations here, the angles $\psi$
and $\omega$ are replaceable by $\lambda$ and $\varpi$ (as we explained in
Sect.~\ref{sec_fit_RV}). The first pair in~(\ref{resang}) corresponds to the 2/1 MMR
between the planets \emph{b} and \emph{c}, the second pair~-- to the same resonance
between \emph{b} and \emph{e}, and the third pair~-- to the 4/1 MMR between \emph{c} and
\emph{e}. All eleven two-planet resonant angles studied by \citet{Rivera10} can be
expressed via the six variables~(\ref{resang}). Namely,
\begin{eqnarray}
\varphi_{cb,c}=-s_{cb,c}, \quad \varphi_{cb,b}=-s_{cb,b}, \quad \varphi_{cb}=s_{cb,b}-s_{cb,c}, \nonumber\\
\varphi_{be,b}=-s_{be,b}, \quad \varphi_{be,e}=-s_{be,e}, \quad \varphi_{be}=s_{be,b}-s_{be,e}, \nonumber\\
\varphi_{ce0}= -3 s_{ce,c}, \quad \varphi_{ce3} = - 3 s_{ce,e}, \quad \varphi_{ce}=s_{ce,c}-s_{ce,e}, \nonumber\\
\varphi_{ce1} = -2 s_{ce,c} - s_{ce,e}, \quad \varphi_{ce2} = -s_{ce,c} - 2 s_{ce,e}.
\label{resvar}
\end{eqnarray}
We prefer to limit ourselves to the minimum possible number of variables. With no loss of
information, we may consider only the behavior of the variables in~(\ref{resang}).

All of the quantities in~(\ref{resang}) are $2\pi$-periodic. It may seem that the
quantities from the last pair of~(\ref{resang}) are $2\pi/3$-periodic, but this is not
strictly true. Adding $2\pi$ to the angle $\psi_c$, for instance, changes $s_{ce,c}$ by
$2\pi/3$ indeed, but the value of $s_{ce,e}$ is changed synchronously. It is better to say
that this pair has a secondary \emph{vectorial} period of $(2\pi/3,2\pi/3)$, in addition
to the usual scalar period of $2\pi$.

\begin{figure}
 \includegraphics[width=0.52\textwidth]{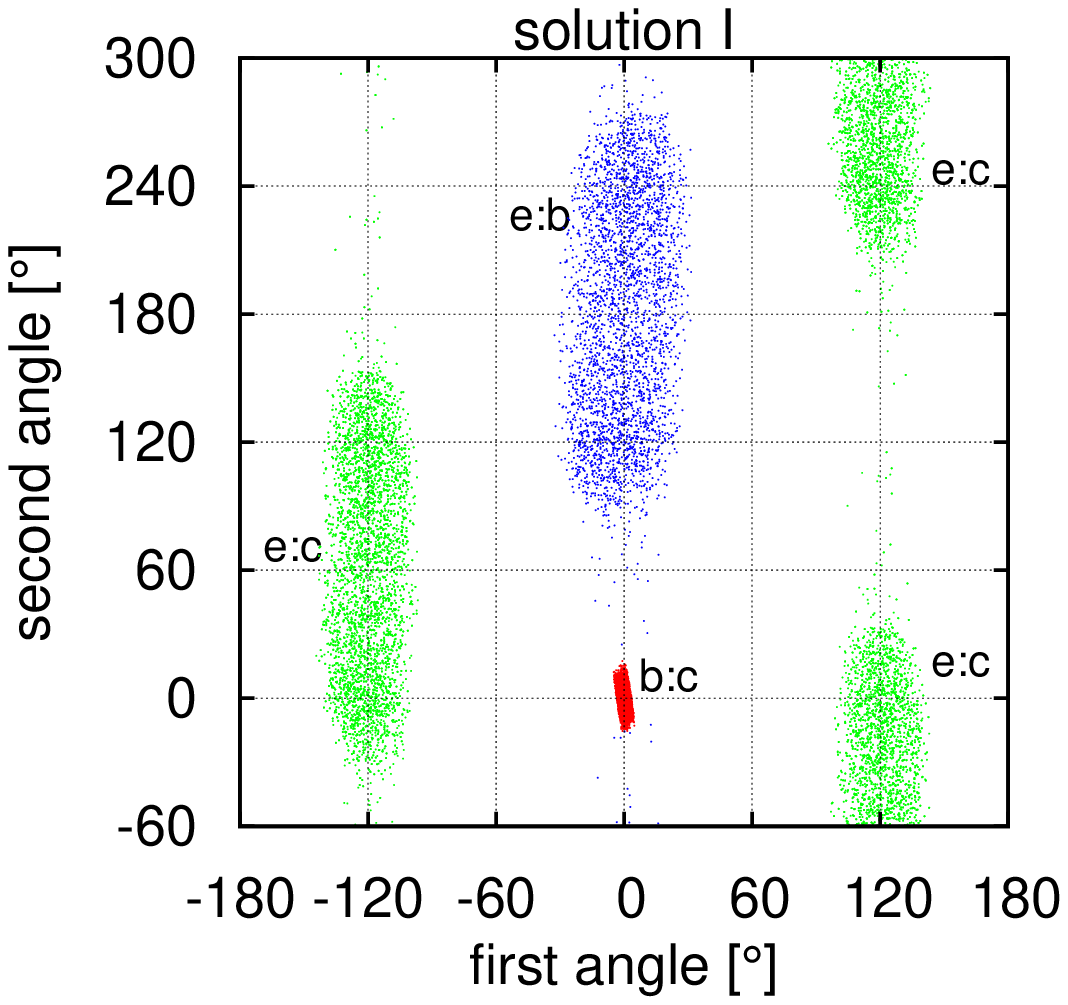}
 \includegraphics[width=0.47\textwidth]{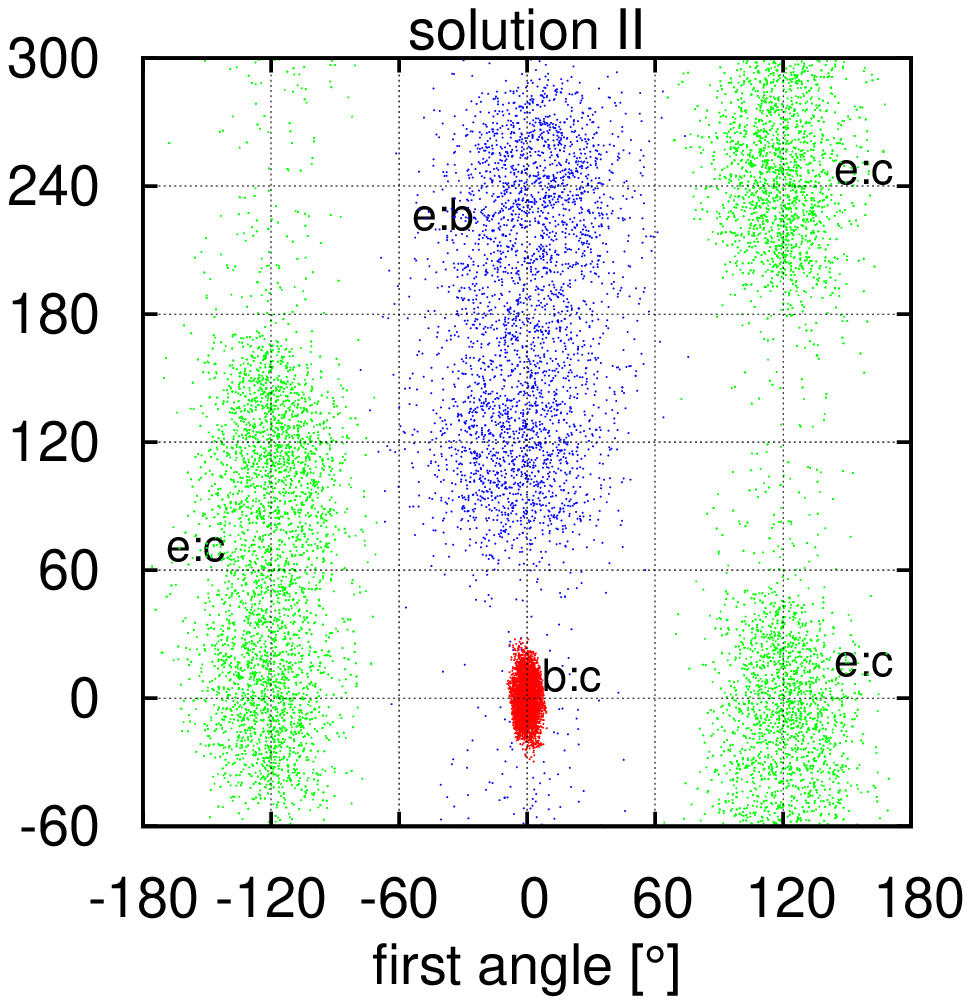}
\caption{Temporal evolution (during the first $10^6$~yrs) of resonant
angles~(\ref{resang}) in the GJ876 planetary system. Each panel shows several dot-filled
domains corresponding to three pairs of the critical angles, referring to the marked
two-planet resonances. Both graphs should also contain one more duplicate ``e:c'' (green)
spot, centered at $(0^\circ,180^\circ)$, but this spot is obscured by the ``e:b'' (blue)
one. We removed this central e:c spot for the clarity of the figure. The third duplicate
e:c spot (second shown) is cut half-and-half in the right part of the both graphs. See
text for the details and discussion.}
\label{fig_libr}
\end{figure}

The evolution of these resonant arguments is illustrated in Fig.~\ref{fig_libr}.
Apparently, all three pairs librate in some limited regions. However, the argument
$s_{be,e}$ actually circulates, systematically avoiding the values $|s_{be,e}|<70^\circ$.
The pair $(s_{ce,b},s_{ce,e})$ behaves in a similar manner: the argument $s_{ce,e}$
circulates, but usually stays in the range $\pm 100^\circ$ around $60^\circ$ (or around
$180^\circ$, or $-60^\circ$, which are equivalent by periodicity). For the solution I,
these results basically agree with the results by \citet{Rivera10}. For the solution II,
the libration ranges become wider, although the picture still remains qualitatively the
same.

We investigated the behavior of the Laplace resonance critical angle, also studied by
\citet{Rivera10}, $s_L = \psi_c - 3\psi_b + 2\psi_e$. For the solution I, this angle
librates between approximately $\pm 40^\circ$, while for the solution II the relevant
amplitude increases roughly twice.

\section{Conclusions}
\label{sec_conc}
The orbit estimations for the GJ876 planetary system are affected by the fine effect of
the correlated data errors, which is rather new to RV planet search surveys. The red RV
noise was responsible, for instance, for the overestimated non-zero values of the planet
GJ876~\emph{d} eccentricity from previous works. Although we still cannot retract as large
values of $e_d$ as $0.15$, the available RV data are consistent with a circular planet
\emph{d} orbit. Another notable red noise effect is a systematic underestimation of the
parameters uncertainties, which occure while we use the traditional white noise model
(compare Tables~\ref{tab_ref},~\ref{tab_refred}). For GJ876, this underestimation is
typically about $10-30\%$, and it is not removed even by rather trusted numerical
simulation method like the bootstrap Monte Carlo. Although the correlated RV noise did not
produce more breaking changes in the GJ876 orbital fit, the very existence of such type of
RV measurements warns us that for other star, especially for those ones where the stellar
jitter dominates in the total RV error budget, the effect of the correlated noise may be
crucial.

\begin{figure}
 \includegraphics[width=\textwidth]{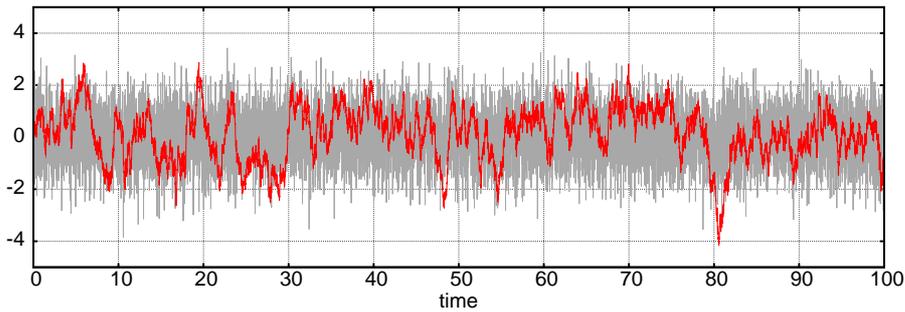}
\caption{Simulated examples of the red and white noise of zero mean and unit variance. The
white noise (background graph) looks like a thick detail-free horizontal band, whereas the
red noise (foreground graph) shows clearly detailed structure, which leaves a very stable
impression that such data contain some mixture of non-random periodic signals. This
oscillating structure gets eventually suppressed at larger time intervals, but
nevertheless is obvious up to tens of the correlation timescales (up to months in case of
GJ876). The red noise was modeled as a Gaussian process with autocorrelation function
$e^{-|\Delta t|}$.}
\label{fig_rednoise}
\end{figure}
A simulated red noise example in Fig.~\ref{fig_rednoise} actually does not look like a
pure noise at all. It looks like a bit noisy mixture of apparently periodic or maybe
non-periodic but non-random components. If we act within the traditional white noise
framework, these false variations may impose a huge misleading effect, and can ultimately
lead to false planet detections.

The planetary perturbations in GJ876 helped us to constrain not only the system
inclination (and therefore the true planet masses), but also the planet \emph{e} orbital
eccentricity ($e_e\lesssim 0.2$), which otherwise could not be limited better then by
$\sim 0.4$. However, on contrary with the inclination, the help with the eccentricity
constraint has a cost. The information constraining the value of $e_e$ comes indirectly
from irregular (probably short-term) perturbations. The irregular nature of the
perturbations made this parameter practically indeterminable inside its admissible region.

We gave improved estimations of all orbital parameters of the system, taking into account
the red noise effect and the bad determinability of the planet \emph{e} eccentric
parameters ($e_e,\omega_e$).

We found the signs of a shallow long-term RV trend of $\sim 0.2$~m/(s$\cdot$yr) in the
data for GJ876. More detailed investigation suggests that the significance of the trend is
tied to the uncertainty in the planet \emph{e} eccentricity. Thus we do not claim that
this trend is real indeed, but we believe it is an issue that should be addressed by
future observations. If this RV trend will be confirmed, it may indicate the existence of
an unseen distant satellite in the system. Its period should exceed $\sim 10-20$~yrs
(observational time span), and the semi-major axis should exceed, consequently, $\sim
3-5$~AU. At the distance from Sun of $4.7$~pc, its sky separation should be $\sim 1''$ or
more, and therefore such object could represent a good target for direct imaging, if it is
large and bright enough. Its mass is poorly constrained, however: it may be as small as
$\sim 0.04 M_{Jup}$ for $P=20$~yrs or arbitrarily larger for longer $P$. This mass scales
as $\propto a^2$ (due to the distance-velocity law for a circular orbit), so for
$a\lesssim 100$~AU that should be a planet rather than a brown dwarf.

Finally, we investigated the system non-coplanarity, and found that although the current
RV data are consistent with the coplanar solution, the actual mutual orbital inclination
between the planets \emph{b} and \emph{c} should not exceed $5^\circ-15^\circ$, depending
on the orientation of the corresponding mutual orbital nodes.

\begin{acknowledgements}
This work was supported by the Russian Academy of Sciences research programme ``Origin and
Evolution of Stars and Galaxies'' and by the President programme of support of leading
scientific schools (grant NSh.3290.2010.2). I am grateful to both referees, who provided
very helpful comments on the manuscript. I am also sincerely thankful to all colleagues
who kindly allowed me to utilize their work computers for weeks, facilitating the enormous
amount of simulations presented in the paper.
\end{acknowledgements}



\appendix
\section{Fitting orbits with correlated RV noise}
\label{sec_fitcorr}
Strictly speaking, we have no goal here to characterize the coloured noise itself; rather,
we want to suppress its influence on the orbital fits. In such case, we may not to care
very much about how well our red noise models describe what really occurs. For example, it
is not very important in practice, to which of the two ``hills'', the low-frequency or the
unit-frequency one, in Fig.~\ref{fig_pow_linfreq} the real correlated noise corresponds,
and which of them is the alias. Both interpretations can almost equally explain the actual
RV data we have, and therefore both types of the noise would eventually lead to similar
effects during the RV curve fitting. Although we have shown in Sect.~\ref{sec_redj} that
the noise non-whiteness significantly exceeds the natural ``white'' random fluctuations in
the data, the latter fluctuations still largely contaminate the noise spectrum. It is
hardly possible that small perturbations of the combined noise power spectrum could
significantly affect our orbital fits, at least in comparison with the usual white-noise
uncertainties. Therefore, we may limit ourselves to only very simple models of the red
noise spectrum, which would only reflect its general monotonic decrease in frequency
(which is actually the only robustly detectable noise spectrum property). The secondary
spectrum excess around the unit frequency will be reproduced automatically due to the
aliasing effect in the time series. We also assume that the red noise component is shared
between both Keck and HARPS data (so they differ only in the white noise parameters). It
is likely that the red noise is just not responsible for the remaining irregular
variations in the HARPS periodogram, since such variations could be expected from the
white noise only.

In addition, we may fix the shape of the autocorrelation function to some mathematically
convenient function, which should only provide a suitably behaving frequency spectrum. In
this paper, we assume that the autocorrelation function of the red error noise
$\epsilon_{\mathrm{red}}(t)$ looks like
\begin{equation}
\mathrm{Corr}(\epsilon_{\mathrm{red}}(t),\epsilon_{\mathrm{red}}(t+\Delta t)) =
 \rho(\Delta t/\tau) = \exp(-|\Delta t|/\tau),
\label{corr}
\end{equation}
where $\tau$ is some unknown parameter characterizing the correlation timescale. According
to the Wiener-Khinchin theorem, the corresponding noise frequency spectrum, $P(f)$,
represents the Fourier transform of~(\ref{corr}) and is equal to
\begin{equation}
P(f) = \int\limits_{-\infty}^{\infty} \rho(\Delta t/\tau) e^{2\pi i f\Delta t}\, d\Delta t =
 \tau \int\limits_{-\infty}^{\infty} \rho(x) e^{2\pi i f\tau x}\, dx = \frac{2\tau}{1+(2\pi f\tau)^2}.
\label{corr-spectrum}
\end{equation}
This function can be in good agreement with what we see in the Keck periodogram, for some
value of the parameter $\tau$. The latter parameter characterizes the width of the
low-frequency band that we see in the periodogram. We also considered some other realistic
shapes of the correlation functions as alternatives: $\rho(x)=\exp(-x^2/2)$,
$\rho(x)=1/(1+x^2)$. They produce a bit different frequency spectra, which are still
generally similar to~(\ref{corr-spectrum}). Note that the white noise is still present in
the data and may be responsible for some constant level in the observed frequency
spectrum. The ratio between the red and white noise contributions is a priori unknown.
Varying $\tau$ and the fraction of the red noise, we may construct a combined model power
spectrum to be in good agreement with the observed one, even for apparently quite
different autocorrelation functions. This means that noise parameters that we will derive
from the RV data may be severely model-dependent and mutually correlated, but also this
means that the red noise effect on the planetary system orbital estimations should be, on
contrary, relatively independent on the noise correlation model (in comparison with the
statistical uncertainties).

The combined \emph{covariance} function $R(t_1,t_2)$ of two different RV measurements,
taken at $t_1$ and $t_2$, is determined by the red noise only:
\begin{equation}
R(t_1,t_2) = \sigma_{\mathrm{red}}^2 \rho((t_2-t_1)/\tau),
\label{cov}
\end{equation}
whereas the variance of a single observations incorporates both white and red components:
\begin{equation}
R(t,t) = \sigma_{\mathrm{white},j}^2 + \sigma_{\mathrm{red}}^2.
\label{var}
\end{equation}
Here the index $j$ refers to the dataset, to which the mentioned observation belongs to.
That is, we assume that the white noise contributions are different for different
datasets, whereas the red noise component is common. Such noise separation could take
place if the red RV noise was actually caused by some activity effects in the stellar
atmosphere.

Since our RV time series are discrete, the covariance function~(\ref{cov},\ref{var})
determines the following $N\times N$ covariance matrix $\tens V$ of all available
RV observations:
\begin{equation}
 \tens V = \tens V_{\mathrm{white}} + \sigma_{\mathrm{red}}^2 \tens R_{\mathrm{red}}(\tau).
\label{covar_matr}
\end{equation}
Here the matrix $\tens V_{\mathrm{white}}$ represents the usual diagonal covariance matrix
of the white part of the noise, and the remaining term is the common red noise component.
The elements of the matrix $\tens R_{\mathrm{red}}$ represent pairwise correlations of the
corresponding RV measurements and are equal to $\rho(\Delta t/\tau)$ (with different
$\Delta t$). When $\sigma_{\mathrm{red}}=0$, we have the usual uncorrelated white noise
model with RV jitter values of $\sigma_{\mathrm{white},j}^2$.

With the noise model~(\ref{covar_matr}), we can no longer use the objective
function~(\ref{lik}) to obtain orbital fits, since this function does not take into
account the red part of the noise. The correct likelihood function should now take into
account the correlations between different RV observations. Since the likelihood function
represent the joint probability density of the observation vector (as inferred by some
fixed values of the model parameters), we must know the joint distribution of the whole
vector of measurements, not just its covariance matrix. Assuming this joint distribution
is multivariate Gaussian (i.e. the noise represents a Gaussian random process), we can
replace~(\ref{lik}) by
\begin{equation}
\ln \tilde{\mathcal L} = -\frac{1}{2} \left( \ln\det\tens V +
 \frac{\vec r^{\mathrm{T}} \tens V^{-1} \vec r}{\gamma} \right) - N\ln\sqrt{2\pi},
\label{lik-corr}
\end{equation}
where $\vec r$ represents the full vector of all RV residuals ($r_{ji}$). We can easily
see that for the white-only noise, when $\tens V$ is diagonal, the
expression~(\ref{lik-corr}) expands to the sum over $N$ measurements in~(\ref{lik}). The
divisor $\gamma$ in~(\ref{lik-corr}) is again needed to reduce the bias in the variance
parameters. The function~(\ref{lik-corr}) involves the following parameters describing the
RV noise structure: the white jitters for different datasets, $\sigma_{\mathrm{white},j}$,
the red jitter, $\sigma_{\mathrm{red}}$, and the correlation timescale of the red jitter,
$\tau$. The maximization of~(\ref{lik-corr}) over the new set of free parameters (the ones
describing the RV noise and the RV curve) yields the necessary best fitting estimations,
which take into account the red noise effect.

\end{document}